\documentclass[10pt]{iopart}
\usepackage{graphicx}
\usepackage{iopams}
\usepackage{hyperref}

\begin{document}

\title{Disordered Hyperuniform Heterogeneous Materials}

\author{Salvatore Torquato$^{1,2,3,4}$}

$^1$  Department of Chemistry,
Princeton University,
Princeton, NJ 08544, USA  \\

$^2$  Department of Physics,
Princeton University,
Princeton, NJ 08544, USA  \\

$^3$  Princeton Institute for the Science and Technology of Materials,
Princeton, NJ 08544, USA \\

$^4$ Program in Applied and Computational Mathematics,
Princeton University,
Princeton, NJ 08544, USA \\

\vspace{1in} \noindent \textbf{Corresponding author contact information}:
\begin{tabbing}
 \hspace{0.25in}
 \= Salvatore Torquato \\
 \>  Tel.: 609-258-3341 \\
 \>  Fax: 609-258-6746 \\
 \>  E-mail: torquato$@$princeton.edu
\end{tabbing}

\noindent \textbf{Short title}: Disordered Hyperuniform Heterogeneous Materials  \newline

%\noindent \textbf{Classification numbers}: ?????

\newpage

\begin{abstract}
% For Physical Biology, should be <200 words.  This is 203 words

Disordered hyperuniform many-body systems are distinguishable states of matter that lie 
between a crystal and liquid: they are like perfect crystals in the way they suppress large-scale density fluctuations and yet are like liquids or glasses 
in that they are statistically isotropic with no Bragg peaks. These systems play a vital role in a number of
fundamental and applied problems: glass formation, jamming, rigidity, photonic and electronic band structure, localization
of waves and excitations, self-organization, fluid dynamics, quantum
systems, and pure mathematics. Much of what we know theoretically about disordered hyperuniform states of matter involves many-particle
systems. In this paper, we derive new rigorous  criteria that disordered hyperuniform
two-phase heterogeneous materials must obey and explore
their consequences. Two-phase heterogeneous media are ubiquitous; examples include composites and porous media, biological media, foams, polymer blends,
granular media, cellular solids,  and colloids. We begin by  obtaining some results that apply to hyperuniform two-phase media in which
one phase is a  sphere packing in $d$-dimensional Euclidean space $\mathbb{R}^d$. 
Among other results, we rigorously establish the requirements for packings of spheres of different sizes
to  be ``multihyperuniform." We then consider hyperuniformity for general two-phase
media in $\mathbb{R}^d$. Here we apply realizability conditions for an autocovariance function and its
associated spectral density of a two-phase
medium, and then incorporate hyperuniformity as a constraint in order to derive new conditions.  We show that some functional forms
can immediately be eliminated from consideration and identify other forms that are allowable. Specific examples and counterexamples are described.
Contact is made with well-known microstructural models (e.g., overlapping spheres and checkerboards) as
well as irregular phase-separation and Turing-type patterns. We also ascertain a family of autocovariance functions
(or spectral densities) that are realizable by disordered hyperuniform two-phase media in any space dimension, 
and present select explicit constructions of realizations. 
These studies  provide insight into the nature of disordered hyperuniformity in the context
of heterogeneous materials and have implications for the design of such novel
amorphous materials.

\end{abstract}

\vspace{2pc}
\maketitle

\section{Introduction}

%biref intro from biol side, mention both structure and properties
%mention the importance of properties in cancer biology
%(especially trapping and diffusion)

The unusual suppression of density fluctuations at large length scales is central 
to the hyperuniformity concept, whose broad importance  for condensed matter physics
and materials science was brought to the fore  only about a decade ago 
in a study that focused on fundamental theoretical aspects, including how it provides
a unified means to classify and categorize crystals, quasicrystals and special
disordered point configurations \cite{To03a}.  Moreover, it was shown that the hyperuniform many-particle systems are poised at a unique type of critical point
in which (normalized) large-scale density fluctuations vanish
such that the direct correlation function of the Ornstein-Zernike relation  is long-ranged \cite{To03a}.
This is to be contrasted
with a standard thermal critical point in which large-scale density fluctuations are
infinitely large and the total correlation function (not the direct correlation function) is long-ranged 
 \cite{Wi65,Ka66,Fi67,Wi74}.

\begin{figure}
\begin{center}
{\includegraphics[  width=2.5in, keepaspectratio,clip=]{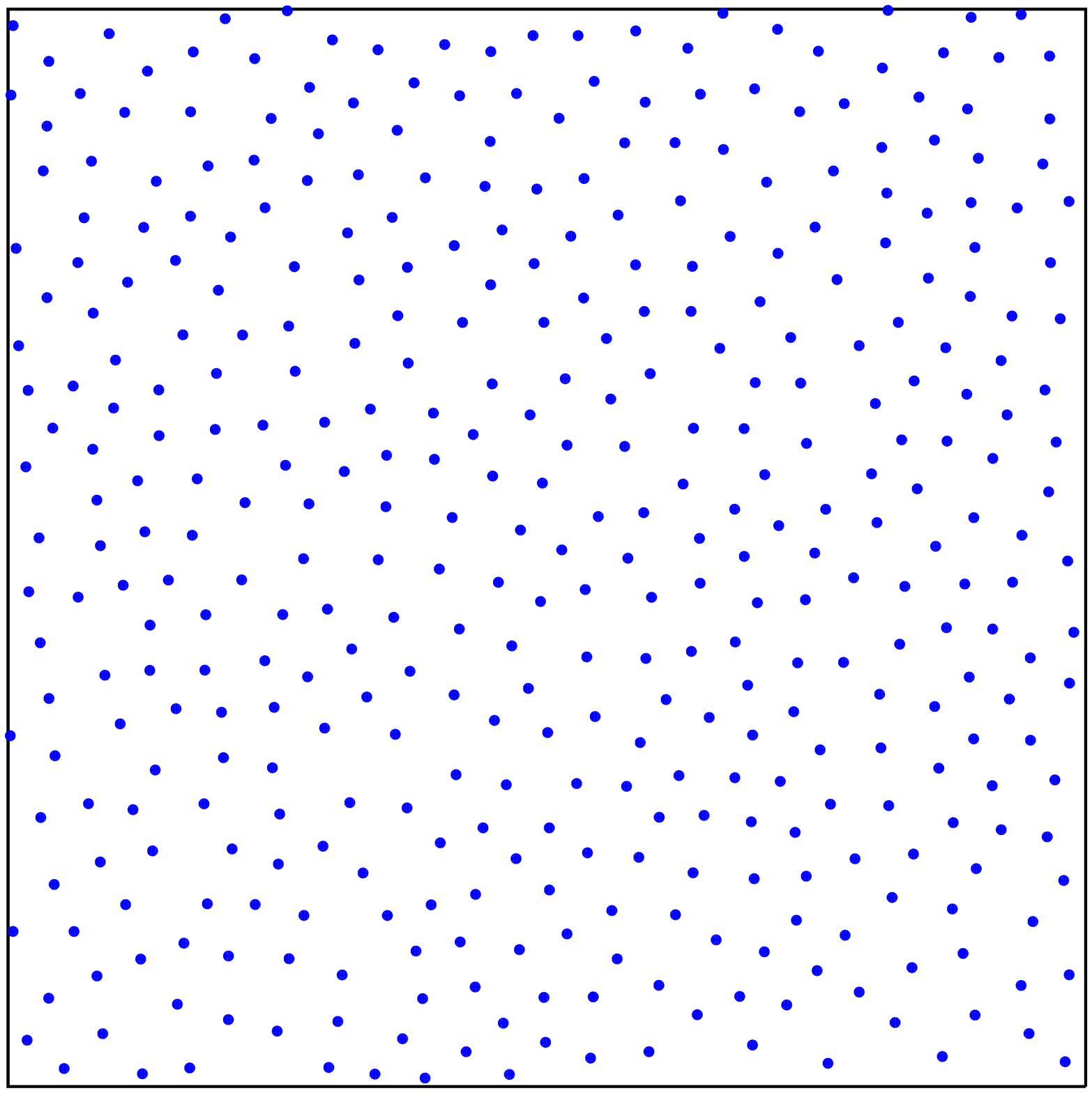}
\includegraphics[  width=2.5in, keepaspectratio,clip=]{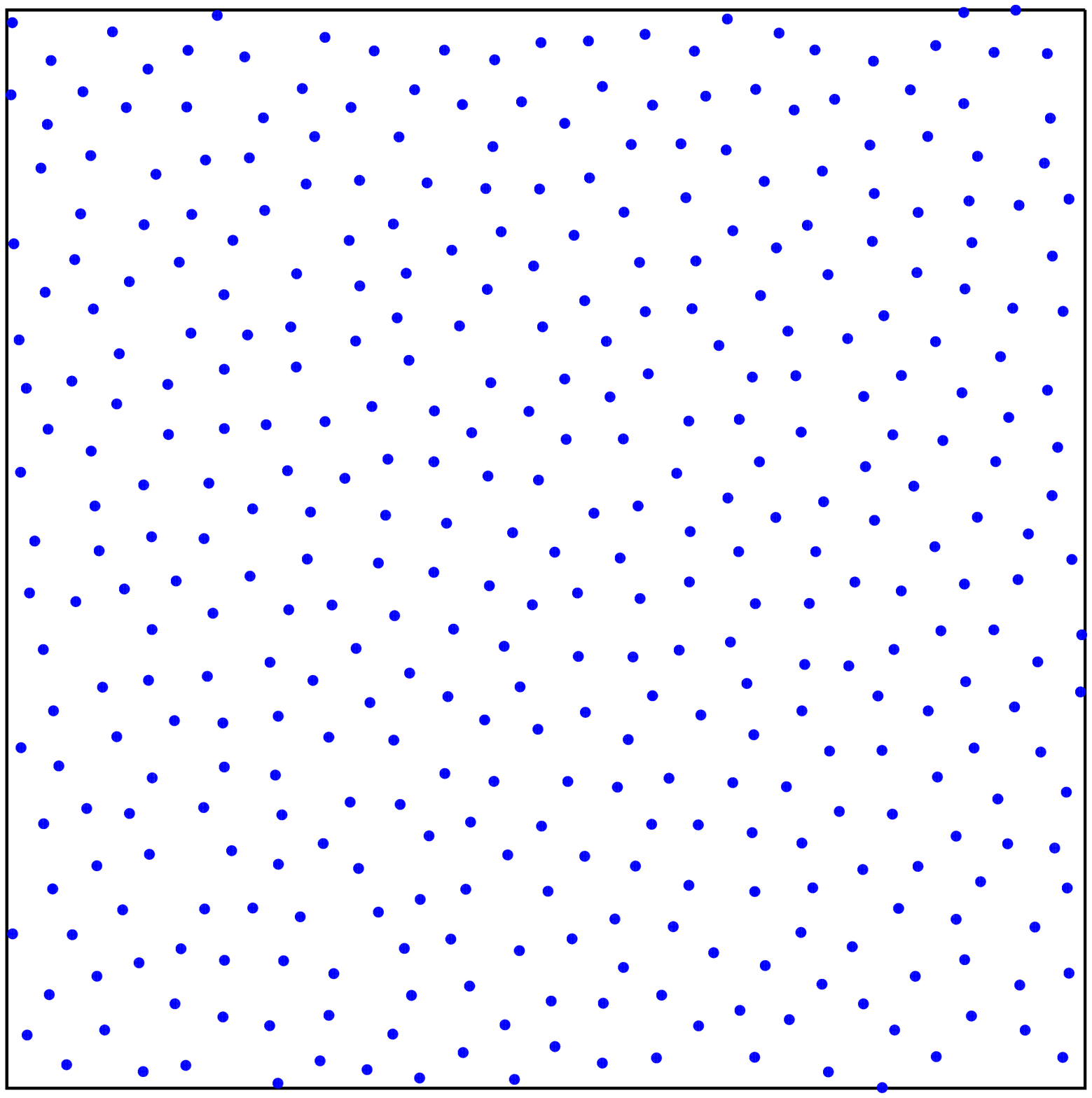}}
\caption{A disordered non-hyperuniform many-particle configuration (left)
and a disordered hyperuniform many-particle configuration (right) \cite{To15}. The latter
is arrived at by very tiny collective displacements of the particles on the left.
These two examples show that it can be very difficult to detect hyperuniformity by eye,
and yet their large-scale structural properties are dramatically different.}
\label{stealthy}
\end{center}
\end{figure}

Roughly speaking, a hyperuniform (or superhomoegeneous \cite{Ga02}) many-particle system in $d$-dimensional Euclidean space
$\mathbb{R}^d$ is one in which (normalized)
density fluctuations are completely suppressed at very large length scales,
implying that the structure factor $S({\bf k})$ tends to zero as the wavenumber $k\equiv |\bf k|$ tends to zero,
i.e.,
\begin{equation}
\lim_{|{\bf k}| \rightarrow 0} S({\bf k}) = 0.
\label{hyper}
\end{equation}
Equivalently, it is one in which the number variance of particles within a
spherical observation window of radius $R$, denoted by $\sigma^2_{_N}(R)$,  grows more slowly than the window volume ($R^d$) in the large-$R$ limit. Typical disordered systems, such as liquids and structural glasses, have the standard volume scaling, that is,  $\sigma^2_{_N}(R) \sim R^d$. By contrast, all perfect crystals and quasicrystals are hyperuniform with the surface-area 
scaling $\sigma^2_{_N}(R)\sim R^{d-1}$. Surprisingly,
there are a special  class of disordered particle configurations, such as the one shown in the right panel of
Fig. \ref{stealthy}, that have the same asymptotic behavior as crystals. There are scalings for the 
number variance other than surface-area growth.  When the structure factor
goes to zero in the limit $|{\bf k}| \rightarrow 0$ with the power-law form
\begin{equation}
S({\bf k}) \sim |{\bf k}|^\alpha,
\label{power}
\end{equation}
where $\alpha >0$, the number variance has the following large-$R$ asymptotic scaling \cite{To03a,Za09,Za11b}:
\begin{equation}
\sigma^2_{_N}(R) \sim\cases{
R^{d-1}, 
     &$\alpha >1$, \cr
 R^{d-1} \ln R, 
     & $\alpha = 1 \qquad (R \rightarrow \infty)$. \cr
 R^{d-\alpha},    
     &  $0 < \alpha < 1$
}
\label{sigma-N-asy}
\end{equation}

Disordered hyperuniform systems can be regarded to be  exotic states
of matter that lie between a crystal and liquid: they are like perfect crystals in the way they suppress large-scale density fluctuations and yet are like liquids or glasses in that they are statistically isotropic with no Bragg peaks. In this sense, they can have a {\it hidden order} (see Fig. \ref{stealthy} for a vivid example) and 
appear to be endowed with novel physical properties, as described below.

\begin{figure}
\begin{center}
\includegraphics*[  width=2.in,clip=keepaspectratio]{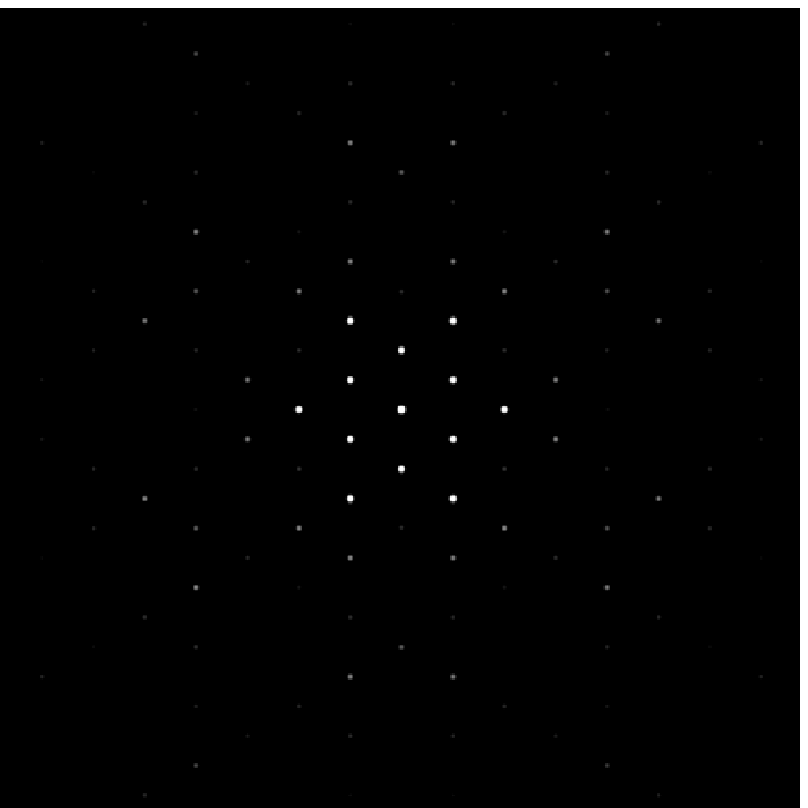}\hspace{0.3in}
 \includegraphics*[  width=2.in,clip=keepaspectratio]{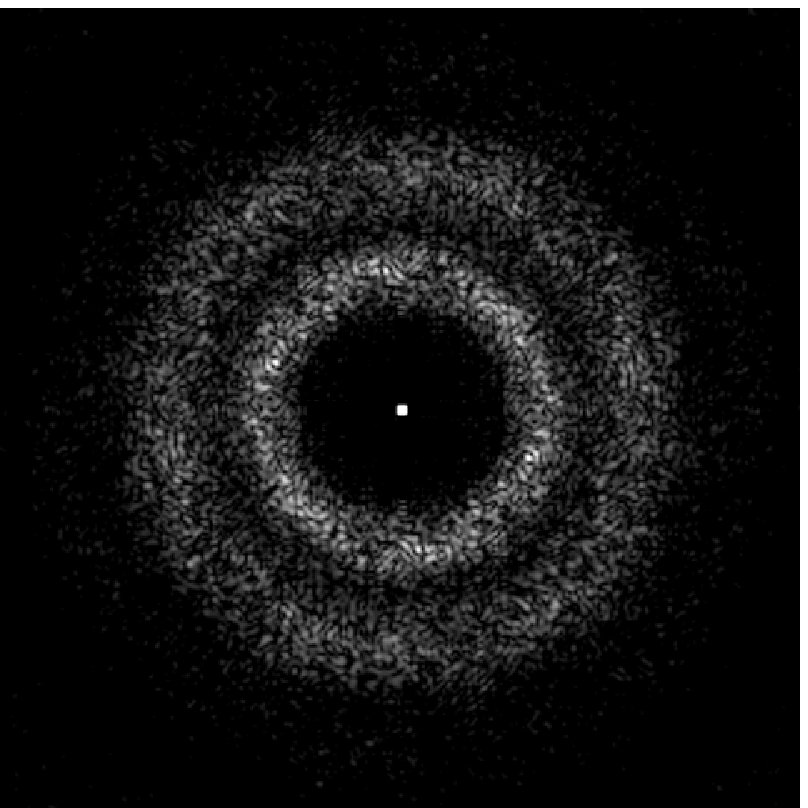}
\caption{Left: Scattering pattern for a crystal. Right: Scattering pattern for a  disordered ``stealthy" hyperuniform material
(defined in Sec. \ref{points}).
Notice that apart from forward scattering, there is a circle around the origin
in which there is no scattering, a highly exotic situation for an amorphous state of matter.}
\label{pattern}
\end{center}
\end{figure}

We knew of only a few examples of {\it disordered} hyperuniform systems 
about a decade ago \cite{To03a,Leb83,Ga02,Do05d}. The importance of the hyperuniformity concept in the context of condensed matter started to become apparent
when it was shown that classical many-particle systems with certain long-ranged
pair potentials could counterintuitively freeze into disordered hyperuniform states
at absolute zero with singular scattering patterns, such as the one shown
in the right panel of Fig. \ref{pattern} \cite{Uc04b,Ba08}. This exotic situation runs counter to our everyday experience where
we expect liquids to freeze into crystal structures (like ice).
Mapping such configurations of particles to network structures, what was previously thought to
be impossible became possible, namely, the first disordered
dielectric networks to have large isotropic photonic band gaps comparable in
size to photonic crystals \cite{Fl09b}.
 We now know that these exotic states of matter can exist as both {\it equilibrium} and {\it nonequilibrium} phases across
space dimensions,
including maximally random jammed particle packings \cite{Za11a,Ji11c,Ch14a},  jammed athermal granular media~\cite{Be11}, jammed thermal colloidal packings~\cite{Ku11,Dr15}, dynamical processes in ultracold atoms~\cite{Le14}, driven nonequilibrium systems \cite{Ja15,He15,We15,Tj15,Di15,Sc15}, avian photoreceptor patterns \cite{Ji14}, geometry of neuronal tracts \cite{Bur15},
certain quantum ground states (both fermionic and bosonic) \cite{To08c,Fe56},  classical disordered
(noncrystalline) ground states \cite{To15,Uc04b,Ba08,Zh15a,Zh15b}.
A variety of groups have recently fabricated disordered hyperuniform materials at the
micro- and nano-scales for various photonic applications \cite{Man13a,Ha13,Man13b},
surface-enhanced Raman spectroscopy \cite{De15}, the
realization of a terahertz quantum cascade laser \cite{Deg15} and self-assembly
of diblock  copolymers \cite{Zi15b}.  Moreover, a computational
study revealed that the electronic bandgap of amorphous silicon
widens as it tends toward a hyperuniform state \cite{He13}. Recent
X-ray scattering  measurements indicate that amorphous-silicon samples
can be made to be nearly hyperuniform   \cite{Xie13}.
Finally, we note that the hyperuniformity concept has suggested new correlation functions
from which one can extract relevant growing length scales as a function of temperature
as a liquid is supercooled below its glass transition temperature \cite{Ma13a},
a problem of intense interest in the glass physics community \cite{Lu07,Be07,Sc07,Ka09,Chand10,Hock12}.

\begin{figure}[bthp]
\centerline{\includegraphics[  width=3.in, keepaspectratio,clip=]{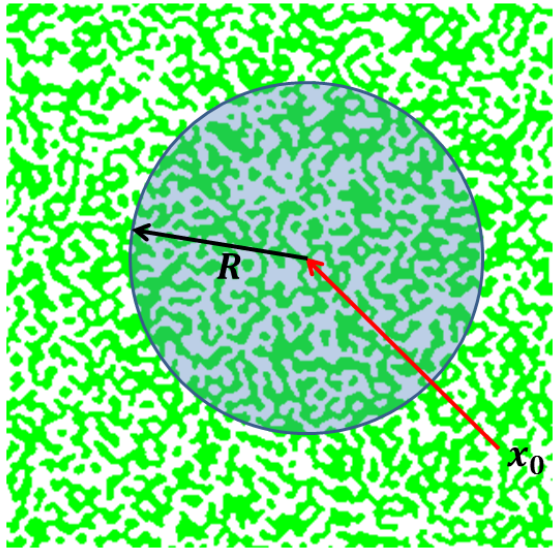}}
\caption{A schematic indicating a circular observation window  of radius $R$ that is centered at
position $\bf x_0$ in a disordered two-phase medium;  one phase is depicted
as a green region and the other phase as a white region. The phase volume fractions  within the window will fluctuate as the window position ${\bf x}_0$ is varied. }
\label{patterns}
\end{figure}

The hyperuniformity concept was generalized to the case of two-phase heterogeneous materials \cite{Za09},
which are ubiquitous; examples include composites and porous media, biological media, foams, polymer blends,
granular media, cellular solids   and colloids \cite{To02a,Sa03}. Here the phase volume fraction
fluctuates within a finite-sized spherical window of radius $R$ (see  Fig. \ref{patterns}) and hence can be characterized by the
volume-fraction variance $\sigma_{_V}^2(R)$. For typical disordered two-phase media, the variance  $\sigma_{_V}^2(R)$ for large $R$ goes to zero 
like $R^{-d}$. However, for hyperuniform disordered two-phase media, $\sigma_{_V}^2(R)$  goes to zero asymptotically more 
rapidly than the inverse of the window volume, i.e., faster than $R^{-d}$, which is equivalent to the following condition
on the spectral density (defined in Sec. \ref{back}):
\begin{eqnarray}
\lim_{|\mathbf{k}|\rightarrow 0}\tilde{\chi}_{_V}(\mathbf{k}) = 0.
\label{hyper-2}
\end{eqnarray}
As in the case of hyperuniform point configurations \cite{To03a,Za09,Za11b},  three different scaling regimes 
when the spectral density goes to zero with the power-law form ${\tilde \chi}_{_V}({\bf k})\sim |{\bf k}|^\alpha$: 
\begin{eqnarray}  
\sigma^2_{_V}(R) \sim \left\{
\begin{array}{lr}
R^{-(d+1)}, \quad \alpha >1\\
R^{-(d+1)} \ln R, \quad \alpha = 1 \qquad (R \rightarrow \infty),\\
R^{-(d+\alpha)}, \quad 0 < \alpha < 1
\end{array}\right.
\label{sigma-V-asy}
\end{eqnarray}
where the exponent $\alpha$ is a positive constant.

Much of our recent theoretical understanding of hyperuniform states of matter is based on many-particle
systems. The purpose of this paper is to delve more deeply into theoretical foundations
of disordered hyperuniform two-phase media by establishing new rigorous  criteria that such systems must obey and exploring
their consequences.

In Sec. \ref{back}, we provide necessary mathematical definitions and background. In Sec. \ref{packing-1},
we derive some results concerning hyperuniformity of two-phase systems in $\mathbb{R}^d$ in which
one phase is a  sphere packing and the spheres generally have different sizes. 
We determine the necessary and sufficient conditions for a sphere packing 
to be stealthy and hyperuniform, and  prove that when each subpacking associated with each component is
hyperuniform, the entire packing is hyperuniform: a property called ``multihyperuniformity"  \cite{Ji14}. 
In Sec. \ref{two}, we consider hyperuniformity for general two-phase
media  that lie outside the special class that are derived from sphere
packings in $d$-dimensional Euclidean space $\mathbb{R}^d$. Here we apply realizability conditions 
for an autocovariance function and its associated spectral density of a two-phase
medium, and then incorporate hyperuniformity as a constraint in order to derive new conditions.   
We demonstrate that some functional forms
can immediately be eliminated from consideration, but also identify other forms that are allowable. 
Specific examples and counterexamples are described, including remarks about well-known microstructural models (e.g., overlapping spheres and checkerboards) as
well as irregular phase-separation and Turing-type patterns. We also ascertain a family of autocovariance functions
that are realizable by disordered hyperuniform two-phase media in arbitrary space dimensions,
In Sec. \ref{con}, we close with some concluding remarks.

\section{Background}
\label{back}

\subsection{Point Configurations}
\label{points}

Consider statistically homogeneous  point configurations in $d$-dimensional Euclidean 
space $\mathbb{R}^d$. The standard pair correlation function $g_2({\bf r})$ is proportional
to the probability density associated with finding pairs of points separated
by the displacement vector $\bf r$, and is normalized in such a way
that it tends to unity in the limit $|{\bf r}| \rightarrow \infty$ in the absence
of long-range order. The {\it total correlation function} $h({\bf r})$ is defined as
\begin{equation}
h({\bf r})=g_2({\bf r})-1.
\label{total}
\end{equation}
The nonnegative structure factor $S(\bf k)$, which is proportional to the scattering intensity, 
is trivially related to the Fourier transform of  $h(\bf r)$:
\begin{equation}
S({\bf k})=1+\rho {\tilde h}({\bf k}).
\label{factor}
\end{equation}
\ref{FT} provides definitions of the $d$-dimensional Fourier transforms that we use in this paper.

The local number variance $\sigma_N^2(R)$ is determined entirely by pair correlations \cite{To03a}:\vspace{-0.1in}
\begin{eqnarray}
\sigma_N^2(R)&=&
 \rho v_1(R)\Big[ 1+\rho\int_{\mathbb{R}^d}  h({\bf r})
\alpha(r;R) d{\bf r}\Big] \nonumber \\
&=&\rho v_1(R)\Big[\frac{1}{(2\pi)^d} \int_{\mathbb{R}^d} S({\bf k}) 
{\tilde \alpha}(k;R) d{\bf k}\Big]  ,
\label{local}
\end{eqnarray}
where  $v_1(R)= \pi^{d/2} R^d/\Gamma(1+d/2)$ is the $d$-dimensional volume of a spherical window, 
$\alpha(r;R)$ is the intersection volume of two identical
hyperspheres of radius $R$ (scaled by the volume of a sphere) whose centers
are separated by a distance $r$, which is known analytically in any space dimension \cite{To03a,To06b},
and ${\tilde \alpha}(k;R)$ is its Fourier transform, which is nonnegative and explicitly given by
\begin{equation}
{\tilde \alpha}(k;R)= 2^d \pi^{d/2} \Gamma(1+d/2)\frac{[J_{d/2}(kR)]^2}{k^d}.
\label{alpha-k}
\end{equation}
Here $J_{\nu}(x)$ is the Bessel function of order $\nu$.

As mentioned earlier, the hyperuniformity property for point configurations is specified by the structure-factor 
condition (\ref{hyper}).
{\it Stealthy} configurations are those in which
the structure factor is exactly zero for a subset of wave vectors, meaning that they completely suppress
single scattering of incident radiation for these wave vectors \cite{Ba08}. 
Stealthy {\it hyperuniform} patterns \cite{To15,Uc04b,Ba08} are a subclass of hyperuniform
systems in which the structure factor is zero for a range
of wave vectors around the origin, i.e.,
\begin{equation}
S({\bf k})= 0 \qquad \mbox{for}\; 0 \le |{\bf k}| \le K,
\end{equation}
\label{stealth}
where $K$ is some positive number. An example of a stealthy disordered scattering pattern
is shown in the right panel of Fig. \ref{pattern}.

\subsection{Two-Phase Media}

A two-phase  random medium is a domain of space $\mathcal{V} \subseteq \mathbb{R}^d$ of volume $V$
that is partitioned into two disjoint regions that make up  $\mathcal{V}$:
a phase 1 region $\mathcal{V}_1$ of volume fraction $\phi_1$ and a phase 2 region $\mathcal{V}_2$ of volume fraction $\phi_2$ \cite{To02a}.

\subsubsection{Two-Point Statistics}

The phase indicator function ${\cal I}^{(i)}({\bf x})$ for a given realization is defined as
\begin{equation}
{\cal I}^{(i)}({\bf x}) = \left\{
{\begin{array}{*{20}c}
{1, \quad\quad {\bf x} \in {\cal V}_i,}\\
{0, \quad\quad {\bf x} \notin {\cal V}_i},
\end{array} }\right.
\label{phase-char}
\end{equation}
\noindent 
The one-point correlation function $S_1^{(i)}({\bf x})= \langle {\cal I}^{(i)}({\bf x}) \rangle$
(where angular brackets indicate an ensemble average) is generally dependent on the position $\bf x$, but 
is a constant for statistically homogeneous media, namely, the phase volume fraction, i.e.,
\begin{equation}
\phi_i = \langle {\cal I}^{(i)}({\bf x}) \rangle,
\end{equation}
such that $\phi_1+\phi_2=1$.
The two-point correlation function is defined as $S^{(i)}_2({\bf x}_1,{\bf x}_2) = \left\langle{{\cal I}^{(i)}({\bf x}_1){\cal I}^{(i)}({\bf x}_2)}\right\rangle$.
 This function is the probability
of finding two points at positions ${\bf x}_1$ and ${\bf x}_2$  in phase $i$.
For statistically homogeneous  media,  the
two-point correlation function will
only depend on the relative displacement vector ${\bf r} \equiv {\bf x}_2-{\bf x}_1$ 
and hence $S_2^{(i)}({\bf x}_1,{\bf x}_2)=S_2^{(i)}({\bf r})$. 
The autocovariance function $\chi_{_V}({\bf r})$ associated with the random variable ${\cal I}^{(i)}({\bf x})$ 
for phase 1 is equal to that for phase 2, i.e.,
\begin{equation}
\label{eq108}
\chi_{_V}({\bf r}) \equiv  S^{(1)}_2({\bf r}) - {\phi_ 1}^2 =  S^{(2)}_2({\bf r}) - {\phi_2}^2.
\end{equation}
At the extreme limits of its argument, $\chi_{_V}$ has the following asymptotic behavior
\begin{equation}
\chi_{_V}({\bf r}=0)=\phi_1\phi_2, \qquad \lim_{|{\bf r}| \rightarrow \infty} \chi_{_V}({\bf r})=0,
\label{limits}
\end{equation}
the latter limit applying when the medium possesses no long-range order. If the medium is statistically homogeneous and isotropic, then the  autocovariance
function ${\chi}_{_V}({\bf r})$ depends only on the magnitude of its argument $r=|\bf r|$,
and hence is a radial function. In such instances, its slope at the origin is directly related 
to the {\it specific surface} $s$ (interface area per unit volume); specifically, we have in any space
dimension $d$, the asymptotic form \cite{To02a},
\begin{equation}
\chi_{_V}({\bf r})= \phi_1\phi_2 - \beta(d) s \;|{\bf r}| + {\cal O}(|{\bf r}|^2),
\label{specific}
\end{equation}
where 
\begin{equation}
\beta(d)= \frac{\Gamma(d/2)}{2\sqrt{\pi} \Gamma((d+1)/2)}.
\label{beta}
\end{equation}

The nonnegative spectral density ${\tilde \chi}_{_V}({\bf k})$, which can be obtained from  scattering experiments \cite{De49,De57},
is  the Fourier transform of $\chi_{_V}({\bf r})$, i.e.,
\begin{equation}
{\tilde \chi}_{_V}({\bf k}) = \int_{\mathbb{R}^d} \chi_{_V}({\bf r}) e^{-i{\bf k \cdot r}} {\rm d} {\bf r} \ge 0, \qquad \mbox{for all} \; {\bf k}.
\label{1}
\end{equation}
 For isotropic media, the spectral density only depends
on $k=|{\bf k}|$ and, as a consequence of  (\ref{specific}), its decay in the large-$k$ limit is controlled
by the exact following power-law form:
\begin{equation}
{\tilde \chi}_{_V}({\bf k}) \sim \frac{\gamma(d)\,s}{k^{d+1}}, \qquad k \rightarrow \infty,
\label{decay}
\end{equation}
where 
\begin{equation}
\gamma(d)=2^d\,\pi^{(d-1)/2} \,\Gamma((d+1)/2)
\end{equation}
is a $d$-dimensional  constant.

The higher-order correlation functions $S_3,S_4,\ldots$
\cite{To02a,To82b,To83a} will not be considered here, but we note that they arise in rigorous bounds
and exact expressions for effective transport \cite{To02a,To85f,Be85a,Be88b,Se89,Gi95a,Mi02,Ph03,To04a}, 
elastic \cite{To02a,Be88b,Gi95a,Mi02,To97b} and electromagnetic \cite{Re08a}
properties of two-phase media.

\subsubsection{Realizability Conditions on Autocovariance Functions of Two-Phase Media}
\label{realize}

A necessary and sufficient condition for the existence of a scalar autocovariance function of a stochastically
continuous homogeneous process is that its spectral function must be a nonnegative 
bounded measure \cite{To02a,Pr81}. However, it is known that for
a two-phase system characterized by the phase indicator function (\ref{phase-char}), the nonnegativity
property of the spectral function [cf. (\ref{1})]
is a necessary but generally not  sufficient
condition for the existence of an autocovariance function $\chi_{_V}({\bf r})$ corresponding
to a  two-phase medium \cite{To02a,To06b,To99c,Ji07,Qu08,Lu15}. 
The autocovariance function must also satisfy other conditions, which are most
conveniently stated in terms of the scaled autocovariance function $f({\bf r})$, which is defined by
\begin{equation}
f({\bf r})\equiv \frac{\chi_{_V}({\bf r})}{\phi_1\phi_2}.
\end{equation}
Comparing this to relation (\ref{limits}), we see that 
\begin{equation}
f({\bf r}=0)=1, \qquad \lim_{|{\bf r}| \rightarrow \infty} f({\bf r})=0.
\label{limits-2}
\end{equation}
We let ${\tilde f}({\bf k})$ denote the Fourier transform of $f({\bf r})$, implying that
\begin{equation}
{\tilde f}({\bf k})=\frac{{\tilde \chi}_{_V}({\bf k})}{\phi_1\phi_2} \ge 0\qquad \mbox{for all} \; {\bf k}.
\end{equation}

Among other conditions, the scaled autocovariance function
must satisfy the following bounds for all $\bf r$:
\begin{equation}
\label{2}
-\min \left[\frac{\phi_1}{\phi_2},\frac{\phi_2}{\phi_1}\right] \le f({\bf r}) \le 1.
\end{equation}
Another necessary condition on $f({\bf r})$ in the case of statistically 
homogeneous and isotropic media, i.e., when $f({\bf r})$ is dependent only 
on the distance $r \equiv |{\bf r}|$, is that its derivative at $r = 0$ is strictly 
negative for all $0<\phi_i<1$:
\begin{equation}
\label{4}
\frac{{\rm d}f}{{\rm d}r} \Bigg|_{r=0} < 0,
\end{equation}
which  is consistent with the fact that slope at $r=0$ is proportional to 
the negative of the specific surface $s$ [cf. (\ref{specific})]. Since 
$f(|{\bf r}|)$ is an even function (i.e., 
$f({\bf r})=f(-{\bf r})$) that is linear in $|{\bf r}|$ at the origin,  it is nonanalytic at the origin.
This is rather a strong restriction because it eliminates
any function that is analytic at the origin (which necessarily implies even
powers of $|{\bf r}|$); for example, it prohibits autocovariance functions
of a Gaussian form [e.g., $\exp(-(r/a)^2)$].
For statistically homogeneous media, another condition is the 
so-called ``triangular inequality'':
\begin{equation}
 f({\bf r}) \ge f({\bf s}) +  f({\bf t}) - 1,
\label{5}
\end{equation}
where ${\bf r} = {\bf t}- {\bf s}$. If the autocovariance function
of a statistically homogeneous and isotropic medium is monotonically 
decreasing, nonnegative and convex (i.e., ${\rm d}^2 f/ {\rm d}^2 r \ge 0$), then 
it satisfies the triangular inequality (\ref{5}). The triangular inequality 
implies several pointwise conditions on $f({\bf r})$. For 
example, for statistically homogeneous and isotropic media, it
implies the condition  (\ref{4})
and  convexity  at the origin:
\begin{equation}
\label{6}
\frac{{\rm d}^2f}{{\rm d}r^2} \Bigg|_{r=0} \ge 0.
\end{equation} 

The triangular inequality is actually a 
special case of the following more general condition:
\begin{equation}
\label{7}
\sum\limits_{i=1}^m\sum\limits_{j=1}^m \varepsilon_i\varepsilon_jf({\bf r}_i-{\bf r}_j)\ge 1,
\end{equation}
where $\varepsilon_i = \pm 1$ ($i = 1,...,m$ and $m$ is odd). Note that by 
choosing $m = 3$; $\varepsilon_1\varepsilon_2 = 1$, $\varepsilon_1\varepsilon_3 = 
\varepsilon_2\varepsilon_3 = -1$, Eq.~(\ref{5}) can be rediscovered. If $m = 3$; 
$\varepsilon_1\varepsilon_2 = \varepsilon_1\varepsilon_3 = \varepsilon_2\varepsilon_3 
= 1$ are chosen instead, another ``triangular inequality'' can be obtained, i.e.,
\begin{equation}
\label{8}
f({\bf r}) \ge  -f({\bf s})- f({\bf t})-1,
\end{equation}
where ${\bf r} = {\bf t}- {\bf s}$. Equation~(\ref{8}) was first 
derived by Quintanilla \cite{Qu08}.
 Equation~(\ref{7}) is a much stronger necessary condition that implies that there
 are other necessary conditions beyond those identified thus far. However, 
Eq.~(\ref{7}) is difficult to check in practice, because it does not have a simple
 spectral analog.

\subsubsection{Local Volume-Fraction Variance and Spectral Density}

It is known that the volume-fraction variance $\sigma_{_V}^2(R)$
within a $d$-dimensional spherical window of radius $R$ can be expressed in terms of the autocovariance function $\chi_{_V}({\bf r})$ \cite{Lu90b}:
\begin{eqnarray}
\sigma_{_V}^2(R) = \frac{1}{v_1(R)} \int_{\mathbb{R}^d} \chi_{_V}(\mathbf{r}) \alpha(r; R) d\mathbf{r},
\label{phi-var-1}
\end{eqnarray}
where 
\begin{equation}
v_1(R) =\frac{\pi^{d/2} R^d}{\Gamma(1+d/2)}
\label{v1}
\end{equation}
is the volume of a $d$-dimensional sphere of radius $R$, and
$\alpha(r;R)$ is the scaled intersection volume, as defined in Eq. (\ref{local}). \footnote{Note that we have 
changed the earlier notation for the volume-fraction variance used in Ref. \cite{Za09} from 
$\sigma^2_{\tau}(R)$ to $\sigma^2_{_V}(R)$ to distinguish it from other variance functions
that have been introduced elsewhere \cite{To16a} to describe generalizations of the
hyperuniformity concept.}
The alternative  Fourier representation of the volume-fraction variance
that is dual to the direct-space representation (\ref{phi-var-1}) is trivially obtained by applying Parseval's theorem to (\ref{phi-var-1})
under the assumption that the spectral density ${\tilde \chi}_{_V}({\bf k})$ [Fourier transform of $\chi_{_V}(\mathbf{r})$] exists:
\begin{eqnarray}
\sigma_{_V}^2(R) = \frac{1}{v_1(R)(2\pi)^d} \int_{\mathbb{R}^d} {\tilde \chi}_{_V}(\mathbf{k}) {\tilde \alpha}(k; R) d\mathbf{k}.
\label{phi-var-2}
\end{eqnarray}
Note that the hyperuniformity condition (\ref{hyper-2}) dictates that the direct-space autocovariance
function $\chi_{_V}({\bf r})$ exhibits both positive and negative correlations such that
its volume integral over all space is exactly zero, i.e.,
\begin{equation}
\int_{\mathbb{R}^d} \chi_{_V}({\bf r}) d{\bf r}=0,
\label{sum-1}
\end{equation}
which can be thought of as a sum rule. The generalization of the hyperuniformity concept to two-phase systems
has been fruitfully applied to characterize a variety of disordered sphere packings \cite{Za11a,Dr15,Za11c, Za11d,Ch15}.

\section{Hyperuniform Sphere Packings}
\label{packing-1}

Here we collect in one place various known results scattered throughout
the literature concerning the autocovariance function
$\chi_{_V}({\bf r})$ and spectral density ${\tilde \chi}_{_V}({\bf k})$ for two-phase
media in $\mathbb{R}^d$ in which one phase is a sphere packing in order to make some remarks  about hyperuniformity and stealthiness.
A particle packing is a configuration of nonoverlapping (i.e., hard)
particles in $\mathbb{R}^d$.
For statistically homogeneous packings of congruent spheres  of radius $a$ in $\mathbb{R}^d$ at number density $\rho$,
the two-point probability function $S_2({\bf r})$ of the particle (sphere) phase is known exactly in terms of the pair correlation function \cite{To02a,To85b},
yielding the autocovariance function as
\begin{eqnarray}
{\chi}_{_V}({\bf r}) &=& \rho\, m(r;a) \otimes m(r;a) +\rho^2 m(r;a) \otimes m(r;a) \otimes h({\bf r}) \nonumber \\
&=& \rho \,v_2^{int}(r;a) +\rho^2 v_2^{int}(r;a) \otimes h({\bf r}),
\label{S2-spheres}
\end{eqnarray}
where 
\begin{equation}
m(r;a) =\Theta(a-r)=\Bigg\{{1, \quad r \le a,\atop{0, \quad r > a,}}
\label{indicator}
\end{equation}
is the sphere indicator function, 
and $v_2^{int}(r;a)=v_1(a)\alpha(r;a)$ is the intersection volume of two spherical windows
of radius $a$ whose centers are separated by a distance $r$, where $v_1(a)$ and $\alpha(r;a)$
are defined as in (\ref{phi-var-1}), and $\otimes$ denotes the convolution of two
functions $F({\bf r})$ and $G({\bf r})$:
\begin{equation}
F({\bf r}) \otimes G({\bf r}) =\int_{\mathbb{R}^d} F({\bf x}) G({\bf r}-{\bf x}) d{\bf x}.
\end{equation}
Fourier transformation of (\ref{S2-spheres}) gives the corresponding spectral
density in terms of the structure factor \cite{Za09,To02a,To85b}:
\begin{eqnarray}
{\tilde \chi}_{_V}({\bf k})&=& \rho \,{\tilde m}^2(k;a)+ \rho^2 {\tilde m}^2(k;a) {\tilde h}({\bf k}) \nonumber \\
&=& \rho\, {\tilde m}^2(k;a) S({\bf k})  \nonumber \\
&=& \phi {\tilde \alpha}(k;a) S({\bf k}) 
\label{chi_V-S}
\end{eqnarray}
where
\begin{equation}
{\tilde \alpha}(k;a)= \frac{1}{v_1(a)} {\tilde m}^2(k;a)= \frac{1}{v_1(a)} \left(\frac{2\pi a}{k}\right)^{d} J_{d/2}^2(ka),
\end{equation}
\begin{equation}
\phi =\rho v_1(a),
\end{equation}
is the {\it packing fraction}, defined to be the fraction of space covered by the nonoverlapping spheres,
and $v_1(a)$ is the volume of a sphere of radius $a$ defined by (\ref{v1}). 

We can bound the volume-fraction variance $\sigma^2_{_V}(R)$ from above in terms of
the number variance $\sigma^2_{_N}(R)$ for some fixed $R$. This is accomplished by
substituting the second line of (\ref{chi_V-S}) into the integral expression (\ref{phi-var-2}),
employing the number-variance relation (\ref{local}) and using the fact that ${\tilde m}(k;a)$ achieves its maximum value of $v_1(a)$
at $k=0$. This leads to the following upper bound:
\begin{equation}
\sigma^2_{_V}(R) \le \left(\frac{a}{R} \right)^{2d} \sigma^2_{_N}(R) \qquad \mbox{for all} \; R,
\label{bound}
\end{equation}
In Ref. \cite{Za09}, the same bound was given, but was derived for the large-$R$ asymptotic limit.
Bound (\ref{bound}) is in fact valid for any $R$.

We now show that the hyperuniformity of a sphere packing in terms 
of volume-fraction fluctuations can only arise 
if the underlying point configuration (determined by the sphere
centers) is itself hyperuniform. Since ${\tilde \alpha}(k;a)$ is analytic at $k=0$,  we have
that in the limit $k \rightarrow 0$,
\begin{equation}
{\tilde \alpha}(k;a)= \frac{\pi^{d/2} R^d}{\Gamma(1+d/2)}
\left[ 1- \frac{(ka)^2}{d+2} + {\cal O}(k^4) \right],
\end{equation}
Because ${\tilde \alpha}(k;a)$ is a positive well-behaved function in the vicinity of   the origin, 
it immediately follows from expression (\ref{chi_V-S}) that if the underlying
point process is hyperuniform, as per the structure-factor condition (\ref{hyper}), then
the spectral density ${\tilde \chi}_{_V}({\bf k})$ inherits the hyperuniformity property (\ref{hyper-2})
only through the structure factor, not ${\tilde \alpha}(k;a)$.
The stealthiness property (no scattering at some finite subset of wave vectors) is a bit more subtle.  We see from relation (\ref{chi_V-S})
that ${\tilde \chi}_{_V}({\bf k})$ is zero at those wave vectors where $S({\bf k})$ is zero as well as
at the zeros
of the function ${\tilde \alpha}(k;a)$, which is determined by the zeros of the Bessel function
$J_{d/2}(ka)$.

To illustrate the utility of these results, we now consider an example where the spectral density
as well as the volume-fraction variance can be calculated exactly for a sphere-packing 
model as density increases up to a maximal value corresponding to hyperuniform state. Specifically, we compute these quantities
for sphere packings corresponding to  a   $g_2$-invariant process 
introduced by Torquato and Stillinger \cite{To03a}. A $g_2$-invariant process
is one in which a chosen nonnegative form for
the pair correlation function $g_2$ remains
invariant over a nonvanishing density range while keeping
all other relevant macroscopic variables fixed \cite{To02b}. The upper
limiting ``terminal'' density is the point above which
the nonnegativity condition on the structure factor
[cf. (\ref{factor})] would be violated. Thus, whenever the structure
factor attains its minimum value of zero at ${\bf k}=0$ at the terminal
or critical density, the system, if realizable, is hyperuniform. 
In Ref. \cite{To03a}, a variety of hyperuniform $g_2$-invariant processes
in which the number variance $\sigma^2_{_N}(R)$ grows like the window surface 
area (i.e., $R^{d-1}$) were exactly  studied in arbitrary space dimensions.
For our purposes, we employ the ``step-function" $g_2$-invariant process, namely, a
$g_2(r)$ that is defined by the unit step function $\Theta(r-D)$, where
$D=2a$ is the sphere diameter. The corresponding  structure factor 
 in the density range $0 \le \rho \le \rho_c$ is given by 
\begin{equation}
S(k)=1-\Gamma(1+d/2) 
\left(\frac{2}{kD}\right)^{d/2}
\left(\frac{\rho}{\rho_c}\right) J_{d/2}(kD),
\label{invariant}
\end{equation}
where  $\rho_c=[2^dv_1(D/2)]^{-1}$ is the terminal density
at which the packing is hyperuniform \cite{To03a}. For $\rho < \rho_c$, the packing 
is not hyperuniform. Substitution of (\ref{invariant}) into relation (\ref{chi_V-S}) yields
the associated spectral density for this model in $d$ dimensions:
\begin{equation}
\hspace{-0.7in}{\tilde \chi}_{_V}({\bf k})=\rho \left(\frac{\pi D}{k}\right)^{d} J_{d/2}^2(kD/2)\Bigg[ 1-\Gamma(1+d/2) 
\left(\frac{2}{kD}\right)^{d/2}
\left(\frac{\rho}{\rho_c}\right) J_{d/2}(kD)\Bigg].
\label{CHI}
\end{equation}
 
\begin{figure}
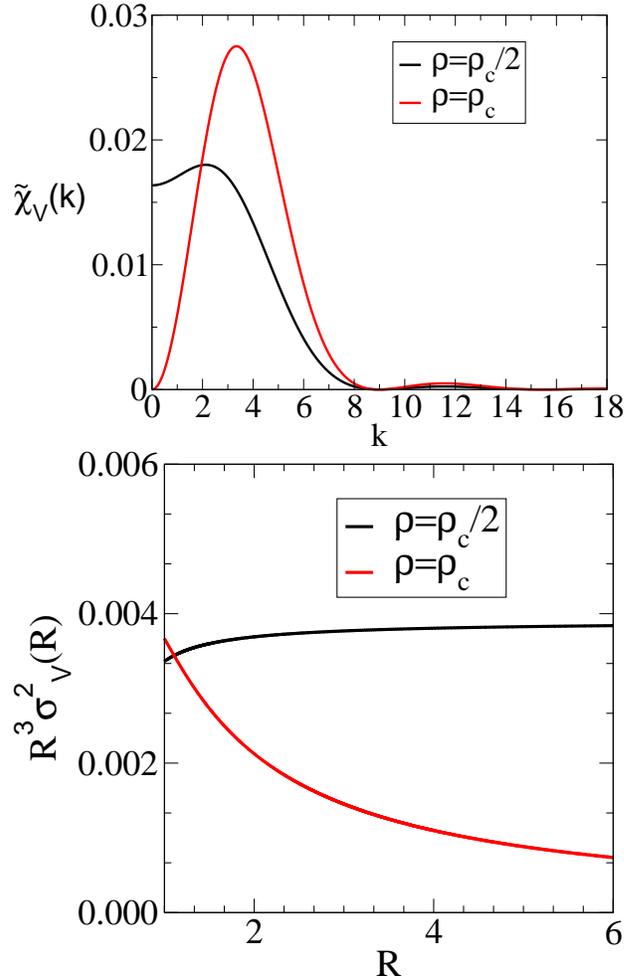

\begin{center}
\includegraphics[  width=3.2in, keepaspectratio,clip=]{fig4a.eps}
 \includegraphics[  width=3.2in, keepaspectratio,clip=]{fig4b.eps}
\caption{Top panel: A hyperuniform spectral density  ${\tilde \chi}_{_V}(k)$ 
versus wavenumber $k$ for sphere packings corresponding to the step-function $g_2$-invariant process in three dimensions 
at two different densities: one at a non-hyperuniform density $\rho=\rho_c/2$ and the other at the hyperuniform terminal density
$\rho_c$, where $\rho_c=\rho=3/(4\pi)$  \cite{To03a}. Bottom panel: The corresponding volume-fraction variance $\sigma^2_{_V}(R)$ 
versus window sphere radius $R$ for the non-hyperuniform and hyperuniform cases.
The diameter of a hard sphere is the unit distance that makes all relevant dimensional variables dimensionless.}
\label{specs}
\end{center}
\end{figure}

The top panel of Fig. \ref{specs} shows the spectral function ${\tilde \chi}_{_V}(k)$ 
for the aforementioned $g_2$-invariant packing process in three dimensions at two different
densities: one at a non-hyperuniform density $\rho=\rho_c/2$ and the other at the hyperuniform terminal density
$\rho_c$, where $\rho_c=\rho=3/(4\pi)$, as obtained from (\ref{CHI}). As noted above, the
degree of hyperuniformity reflected in ${\tilde \chi}_{_V}(k)$ is inherited from the properties
of the structure factor. Note that value of the spectral density at the origin for $\rho=\rho_c/2$
would monotonically decrease as the density increases up to the terminal density at which
point it is exactly zero.
The bottom panel of this figures depicts the associated local volume-fraction
variance $\sigma^2_{_V}(R)$   multiplied by $R^3$ for these two packings, as obtained
from relation (\ref{phi-var-2}). Observe that because $\sigma^2_{_V}(R)$ for  the non-hyperuniform curve  
decays like $R^{-3}$ for large $R$,  the product $\sigma^2_{_V}(R) R^3$ asymptotes to a constant value.
By contrast, the product  $\sigma^2_{_V}(R) R^3$ for $\rho=\rho_c$ decays like $R^{-1}$ for large $R$,
as it should for this three-dimensional hyperuniform
two-phase system.

The aforementioned  results for the pair statistics in both direct and Fourier spaces 
for identical spheres have been generalized to the case of impenetrable spheres
with a continuous or discrete size distribution at overall number density $\rho$ \cite{To02a,Lu91}.
We collect these results in \ref{size} in order and  prove there that when each subpacking associated with each component is
hyperuniform, the entire packing is hyperuniform,
what has been termed {\it multihyperuniformity} \cite{Ji14}.
 It is important to note that examining the structure factor $S({\bf k})$ of the point
configurations derived from the centers of  spheres with a polydispersity in size could lead one to incorrectly
conclude that the packings were not hyperuniform.  It has been demonstrated \cite{Za11a,Za11c,Za11d}
that the proper means of investigating hyperuniformity in this case is through a packing's spectral
density ${\tilde \chi}_{_V}({\bf k})$. This has also been confirmed in experimental studies
of maximally random jammed packings of colloidal spherical particles with a size distribution  \cite{Dr15}.

\section{Hyperuniformity Conditions for a General Class of Two-Phase Media}
\label{two}

Our interest here is to elucidate our understanding of hyperuniformity in general two-phase
media  that lie outside the special class that are derived from sphere
packings, as per the previous section. This is accomplished by applying the
realizability conditions for an autocovariance function of a two-phase
medium that is also hyperuniform. We show that some functional forms
can immediately be eliminated from consideration and that other forms
are allowable. Specific examples and counterexamples are described.
We note that it trivially follows from (\ref{sum-1}) that the scaled autocovariance
$f({\bf r})$ obeys the sum rule
\begin{equation}
\int_{\mathbb{R}^d} f({\bf r}) d{\bf r}=0.
\label{sum-2}
\end{equation}
When $f({\bf r})$ is a function of the modulus $r=|\bf r|$, this sum rule
reduces to the following one-dimensional integral condition:
\begin{equation}
\int_{0}^{\infty} r^{d-1} f(r) dr=0.
\label{sum-3}
\end{equation}

\subsection{Monotonic Autocovariance Functions}

\begin{figure}
\begin{center}
{\includegraphics[  width=2.3in, keepaspectratio,clip=]{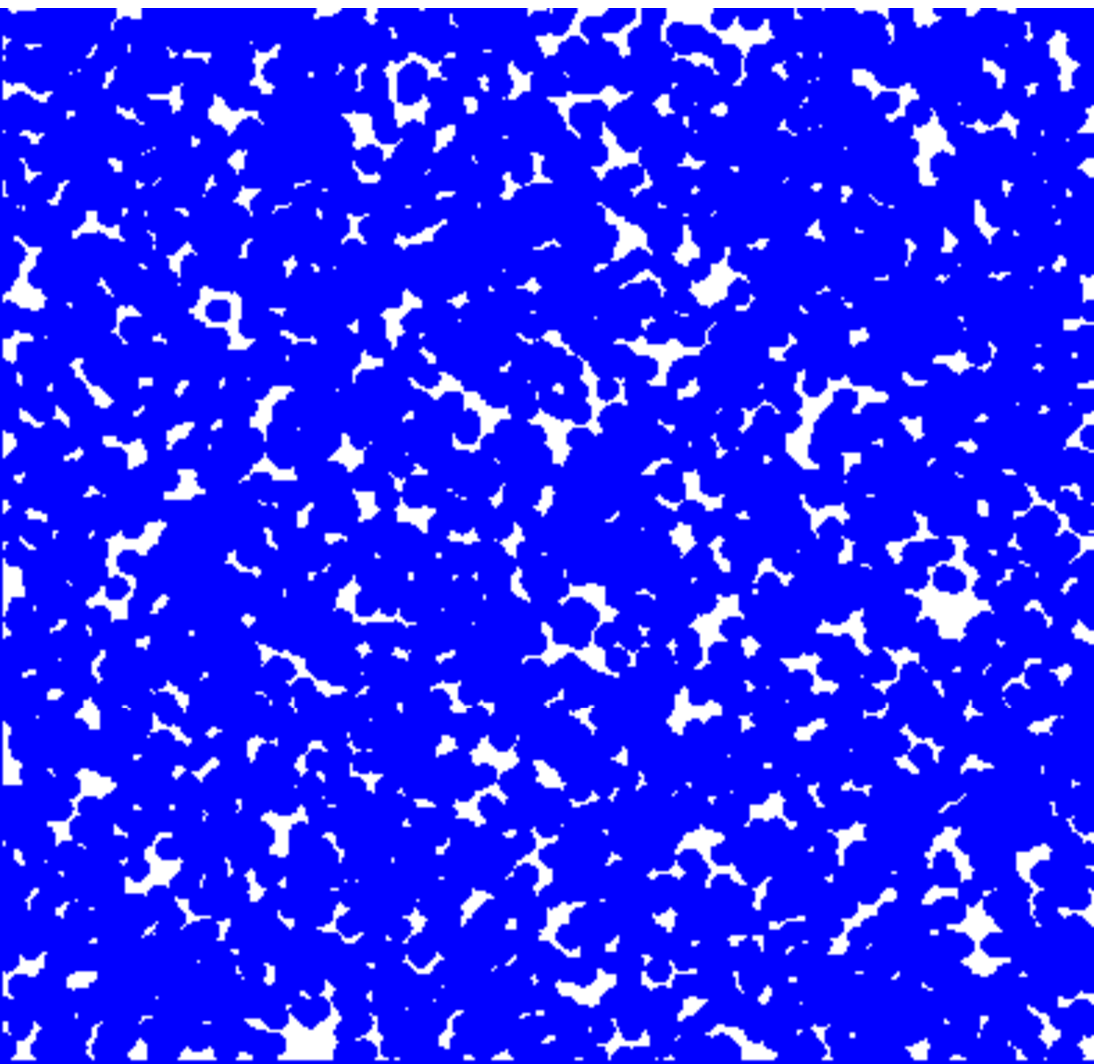}\hspace{0.2in}
\includegraphics[  width=2.2in, keepaspectratio,clip=]{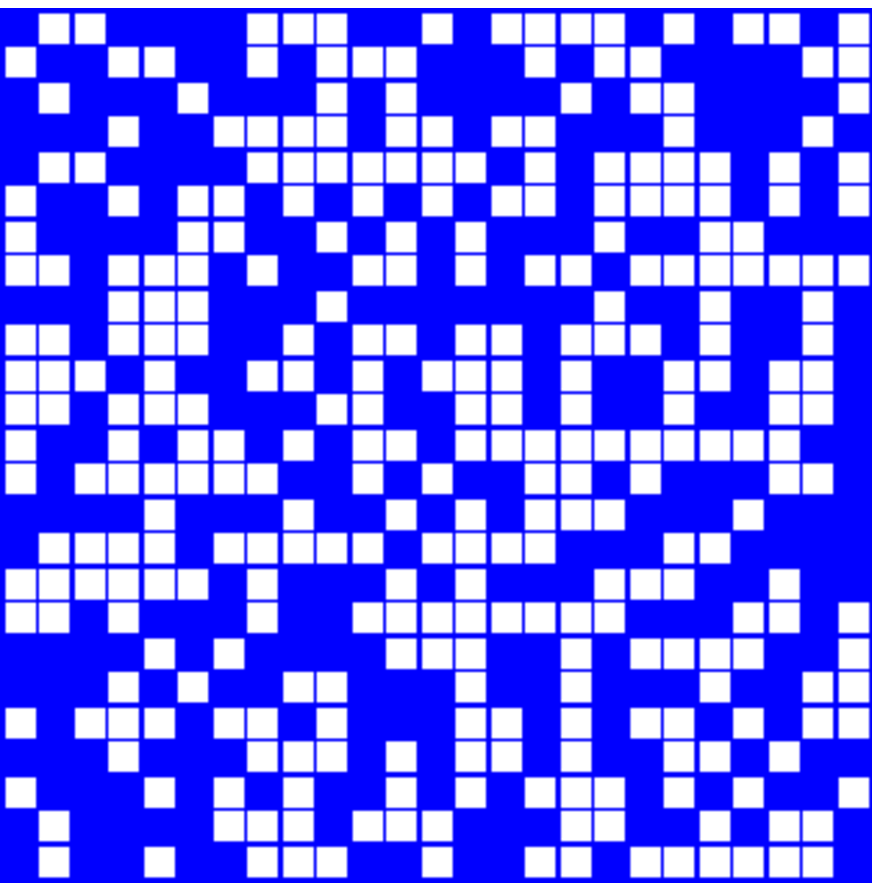}}
\caption{Realizations of overlapping circular disks at $\phi_2=0.885$ (left) and 
of a random checkerboard at $\phi_2=0.5$.}
\label{over}
\end{center}
\end{figure}

To begin, it is instructive to illustrate
the capacity of the sum rule  (\ref{sum-2}) to eliminate an enormous set
of two-phase structures from the hyperuniform class. First, we make the
simple observation that any two-phase medium with a scaled autocovariance 
function $f({\bf r})$ that {\it monotonically decreases} from its maximum
value of unity at the origin to its long-range value, such as the well-known overlapping-sphere
and symmetric-cell  models \cite{To83b,To02a}, cannot be hyperuniform
at any positive volume fraction, since the sum rule  (\ref{sum-2}) 
requires that the autocovariance function $\chi_{_V}({\bf r})$ possess both positive and negative values such that
its volume integral over all space be zero. The overlapping-sphere model in $\mathbb{R}^d$ consists
of the union of spheres that circumscribe the points generated
from a Poisson distribution. The symmetric-cell model is derived
from a tessellation of space into ``cells" with cells being randomly designated
as phase 1 and phase 2 with probability $\phi_1$ and $\phi_2$, respectively.
Figure \ref{over} shows two-dimensional realizations of each of these models. We note that while
these are idealized models, there are many real two-phase systems (e.g.,
sandstones and ceramic-metal composites) that have similar monotonic
autocovariance functions \cite{To02a,Co96} and hence can be immediately ruled out as hyperuniform structures. 
Moreover, it is noteworthy that there is a huge class of two-phase systems that exhibit
strong positive and negative pair correlations at small pair distances
(e.g., equilibrium and nonequilibrium distributions of nonoverlapping
particles) that nonetheless are not hyperuniform by virtue of the fact that their autocovariance functions
violate the sum rule (\ref{sum-2}) \cite{To85a,To85b,To02a}.

\subsection{Remarks About Phase-Separation and Turing Patterns}

\begin{figure}
\begin{center}
\includegraphics[  width=3.5in, keepaspectratio,clip=]{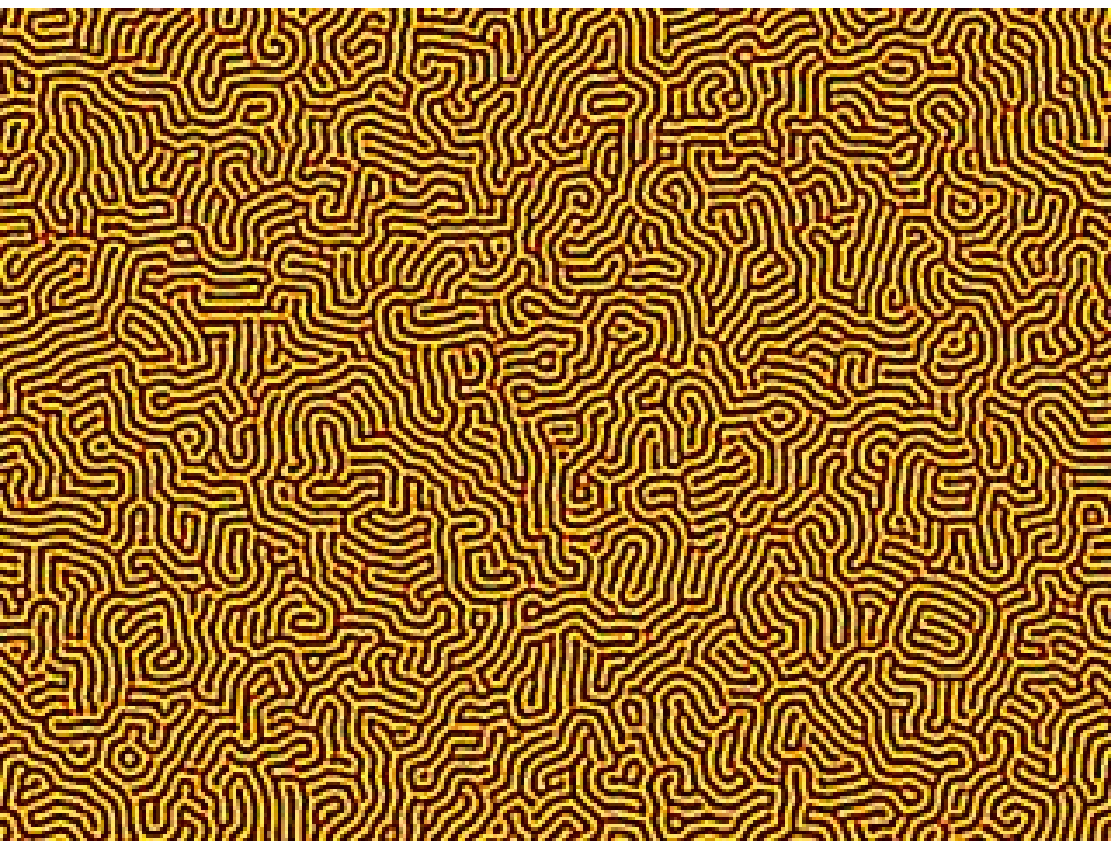}\vspace{0.2in}
\includegraphics[  width=3.5in, keepaspectratio,clip=]{fig6b.eps}
\caption{ Top left panel: Image of a Turing pattern with a labyrinth-like structure \cite{Wiki}.
Bottom panel: The autocovariance function $\chi_{_V}({\bf r})$ associated with the thresholded
version of the Turing image (with $\phi_1 \approx \phi_2 =1/2$), showing strong short-range order,
including anti-correlations (negative values).
The unit of distance  is the mean chord length of the ``yellow" phase \cite{To02a}, which is roughly equal to the characteristic
width of the ``channels." }.
\label{turing1}
\end{center}
\end{figure}

\begin{figure}
\begin{center}
\includegraphics[  width=3.5in, keepaspectratio,clip=]{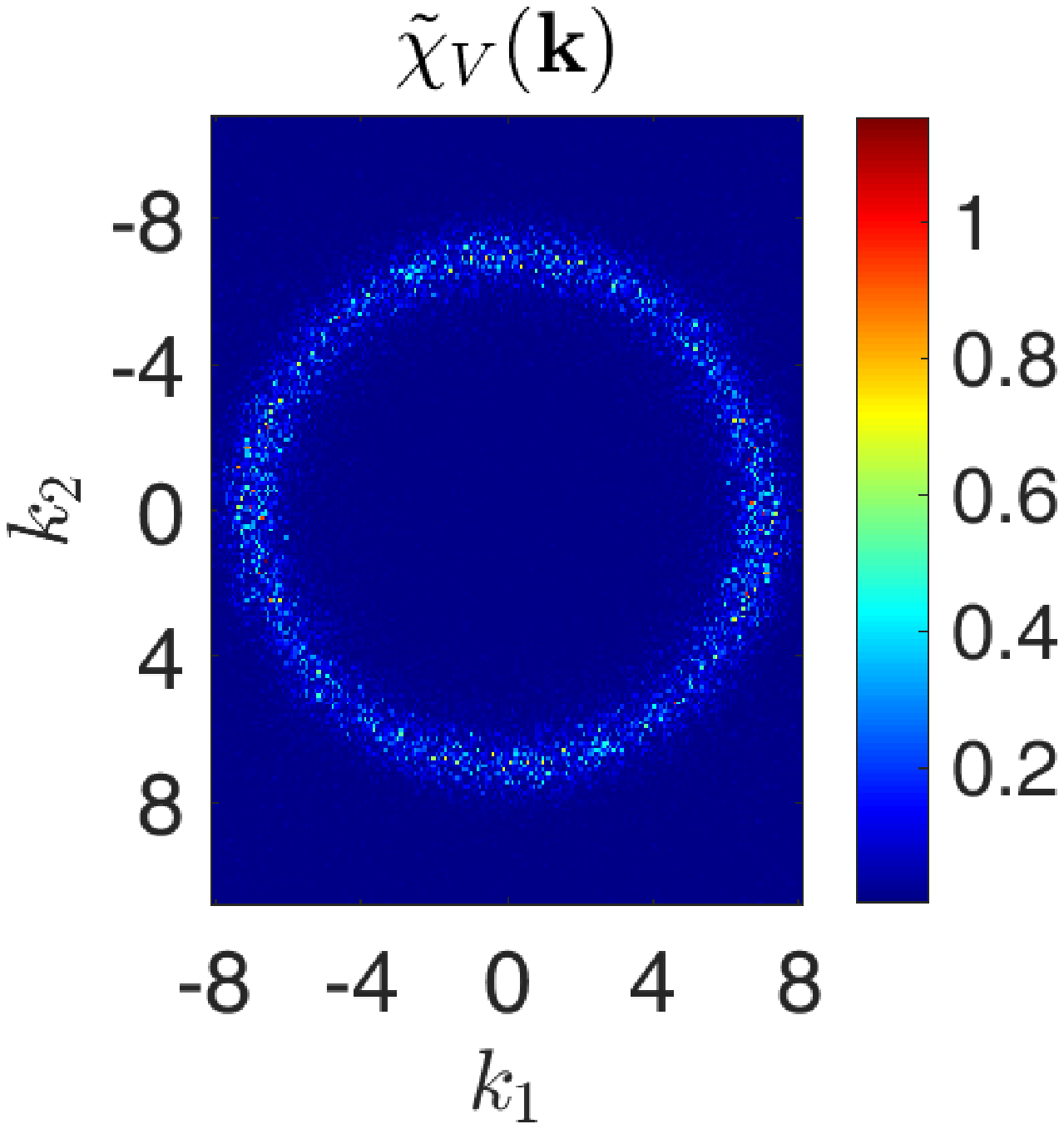}\vspace{0.2in}
\includegraphics[  width=3.5in, keepaspectratio,clip=]{fig7b.eps}
\caption{ Top panel:  Scattering pattern 
as obtained from the spectral density ${\tilde \chi}_{_V}({\bf k})$ associated with the thresholded
version of the Turing image shown in Fig. \ref{turing1}.     Bottom panel: Angular-averaged spectral
density, ${\tilde \chi}_{_V}(k)$, obtained from the 2D scattering pattern shown in the top panel. The unit of distance used in both spectral plots is the mean chord length 
of the ``yellow" phase \cite{To02a}.}
\label{turing2}
\end{center}
\end{figure}

There are a  variety of interesting spatial patterns that arise
in biological and chemical systems that result from a  competition
between different pattern instabilities with selected wavelengths.
Such phenomena have been theoretically described by, for example,
Cahn-Hilliard equations \cite{Ca58} and Swift-Hohenberg equations \cite{Sw77}, whose solutions
can lead to irregular phase-separation and Turing patterns
with a well-defined characteristic wavelength. Thus, it is plausible that
binarized (two-phase) patterns
obtained by thresholding such scalar fields  might be hyperuniform or even stealthy and hyperuniform. An example
of a Turing pattern with an irregular labyrinth-like structure \cite{Wiki} is shown
in Fig. \ref{turing1}. The distance between adjacent  ``channels" of the labyrinth-type pattern is a physical
display of the wavelength (or wavenumber) that has been selected, which
is roughly equal to the mean chord length $\ell_C$ \cite{To02a}. Also, depicted in this
figure is the autocovariance function $\chi_{V}({\bf r})$ associated with the thresholded binarized (two-phase)
version of the Turing image. This function exhibits strong  positive as well as negative correlations
at short distances.
The top panel of Figure \ref{turing2} shows the spectral density ${\tilde \chi}_{_V}({\bf k})$ 
obtained from the thresholded image. Note that it exhibits a well-defined
annulus in which scattering intensity is enhanced relative to that in the region outside
this annulus,  which is radially centered  at $k\, \ell_C \approx 7$. The bottom panel
of Fig. \ref{turing2} shows the angular-averaged spectral density from which we conclude
that the thresholded Turing pattern is neither stealthy nor hyperuniform.

While this outcome does not mean that thresholded Turing-type patterns
can never be hyperuniform, it does lead to the following question: Are there disordered stealthy and hyperuniform two-phase systems 
with spectral densities in which scattering is concentrated within some relatively thin annulus 
defined by a small range of wavenumbers away from the origin?
To answer this question, we consider the following hypothetical, idealized scaled 
spectral functions to see if they can fall within this possible stealthy and hyperuniform class:
\begin{equation}
{\tilde f}_A({\bf k})= c_A(d) \delta(k-K)
\label{A}
\end{equation}
and
\begin{equation}
{\tilde f}_B({\bf k}) =\Bigg\{{c_B(d), \quad K_1\le  k \le K_2,\atop{\hspace{-0.1in} 0, \quad\mbox{otherwise},}}
\label{B}
\end{equation}
where $\delta(k)$ is a radial Dirac delta function is $d$-dimensional
Fourier space, $K_2 > K_1$, 
\begin{equation}
c_A(d)= \frac{2^{d-1}\, \pi^{d/2}\, \Gamma(d/2)}{K^{d-1}}
\end{equation}
and 
\begin{equation}
c_B(d)= (2 \pi)^{d/2} \,\Gamma(d/2+1) \frac{(K_1^d -K_2^d)}{K_1^d K_2^d}.
\end{equation}
Using the results of \ref{FT}, the corresponding
hypothetical scaled autocovariance function, which obeys the exact limiting conditions (\ref{limits-2}), are given by
\begin{equation}
f_A({\bf r})=
\left(\frac{2}{K r}\right)^{d/2-1} \, \Gamma(d/2) \,J_{d/2-1}(K r),
\label{AA}
\end{equation}
and
\begin{equation}
f_B({\bf r})=F(r;K_2)-F(r;K_1),
\label{BB}
\end{equation}
where
\begin{equation}
F(r;K)= c_B(d) \left(\frac{1}{2\pi K r}\right)^{d/2} K^d J_{d/2}(K r).
\end{equation}

Using the results of \ref{FT}, we can expand the aforementioned putative autocovariance functions
about $r=0$ to yield 
\begin{equation}
f_A({\bf r})= 1- C_A(d) r^2  +{\cal O}(r^4)
\end{equation}
and
\begin{equation}
f_B({\bf r})= 1- C_B(d) r^2  +{\cal O}(r^4),
\end{equation}
where $C_A(d)$ and $C_B(d)$ are positive $d$-dimensional constants. It immediately follows 
the autocovariance functions (\ref{AA}) and (\ref{BB}) cannot be realizable by two-phase media since
such systems would have a vanishing specific surface $s$, i.e., the small-$r$ expansion
of a valid autocovariance function must be nonanalytic at the origin such
that the slope is strictly negative [cf. (\ref{realize}]. This strongly suggests
that scattering patterns in which power is concentrated within some concentric
ring of the origin cannot be derived from a two-phase medium. Indeed, any function
that is analytic at the origin cannot be an autocovariance function that
corresponds to a two-phase medium.

\subsection{General Considerations}

A general formalism has been proposed  that enables the functional form 
of a realizable autocovariance function to be expressed by a set of chosen 
realizable basis functions \cite{Ji07}.   For our limited purposes
in this paper, we will make use of only some of these results.
It is known that convex combinations of a set of realizable scaled
autocovariance functions $f_1({\bf r}),f_2({\bf r}),\cdots,f_m({\bf r})$ is itself
a realizable autocovariance function $f({\bf r})$ \cite{Ji07}, i.e.,
\begin{equation}
f({\bf r})= \sum_{i=1}^m \alpha_i f_i({\bf r}),
\label{basis}
\end{equation}
where $0 \le \alpha_i \le 1$ ($i=1,2,\ldots,m$) such that $\sum_{i=1}^m \alpha_i=1$.

In what follows, we focus on basis functions that could correspond to statistically homogeneous and isotropic two-phase
media. A  simple choice   is  the radial exponential function:
\begin{equation}
f_1({\bf r})=\exp(-r/a),
\label{a}
\end{equation}
which is itself a realizable autocovariance function for all positive and finite $a$ \cite{Ji07}.
For reasons discussed at the beginning of this section, 
the monotonicity of $f_1$ precludes it from ever corresponding to a hyperuniform two-phase
system. It has been shown that a linear combination of $f_1$ and  the  basis function
\begin{equation}
f_2({\bf r}) =  \exp(-r/b)\cos(q r + \theta)
\label{b}
\end{equation}
may be realizable for some parameters, but whether such a linear combination can ever correspond 
to a disordered hyperuniform two-phase system has heretofore not been studied. Here $b$ can be thought of as a characteristic
correlation length and $q$ determines the characteristic wavelength  associated with the oscillations. 

Therefore, we explore here whether a disordered hyperuniform two-phase can have  an autocovariance function of the form
\begin{equation}
f({\bf r})= \alpha_1 \exp(-r/a) +\alpha_2  \exp(-r/b)\cos(q r + \theta),
\label{case}
\end{equation}
where $\alpha_1+\alpha_2=1$. For simplicity, we examine two special cases. First, we consider the instance
in which $\alpha_1=0$, $\alpha_2=1$ and $\theta=0$, i.e.,
\begin{equation}
f({\bf r})= \exp(-r/b)\cos(q r).
\label{I}
\end{equation}
Notice that the specific surface corresponding to (\ref{I}) is given by $s=\beta(d)/(b\phi_1\phi_2)$,
where $\beta(d)$ is the $d$-dimensional constant specified in  (\ref{beta}).
The hyperuniformity sum rule (\ref{sum-3}) provides conditions on the parameters $b$ and $q$,
which will depend on the dimension. For example, for $d=1$, we immediately conclude that
(\ref{I}) can never correspond to a hyperuniform medium because (\ref{sum-3}) cannot
be satisfied. On the other hand, for $d=2$ and
$d=3$, hyperuniformity requires that $(qb)^2=1$ and $3(qb)^2=1$, respectively, implying 
that the autocovariance function for a hyperuniform system in a particular dimension generally does not correspond to a hyperuniform system in another dimension. Moreover,
these are only necessary conditions on the parameters $b$ and $q$ for the existence of a hyperuniform two-phase medium and one must still
check whether the known realizability conditions for two-phase media (described Sec. \ref{realize}) are satisfied. As it turns out,
all of these realizability conditions are satisfied, including the nonnegativity
of the spectral density [cf. \ref{1}]. Under these hyperuniform restrictions, the small-$k$ behavior
of the spectral density ${\tilde f}({\bf k})\equiv {\tilde \chi}_{_V}({\bf k})/(\phi_1\phi_2)$ associated
with (\ref{I})  for $d=2$ and $d=3$ are given
respectively by
\begin{equation}
\hspace{-1in}\frac{{\tilde f}({\bf k})}{b^2}= \frac{3}{4\pi} (kb)^2-\frac{35}{128 \pi} (kb)^6 +\frac{693}{8192 \pi} (kb)^{10} +{\cal O}(k^{14}), \quad [(qb)^2=1]
\label{exp-2}
\end{equation}
and
\begin{equation}
\hspace{-1in}\frac{{\tilde f}({\bf k})}{b^3}= \frac{27}{4\pi} (kb)^2-\frac{243}{32 \pi} (kb)^4 +\frac{3645}{512 \pi} (kb)^8- \frac{6561}{1024\pi} (kb)^{10} +{\cal O}(k^{14}), \quad [3(qb)^2=1],
\label{exp-3}
\end{equation}
where we have made use of the small-argument asymptotic expansion of the Bessel function $J_{\nu}(x)$ given in \ref{FT}.
Notice also that the hyperuniformity constraint prohibits multiple powers of four in the two-dimensional expansion (\ref{exp-2}) 
and multiple powers of six in the three-dimensional expansion (\ref{exp-3}). In the opposite asymptotic large-$k$ limit,
we respectively have for $d=2$ and $d=3$
\begin{equation}
\frac{{\tilde f}({\bf k})}{b^2} \sim \frac{2\pi}{(kb)^3}, \qquad k \rightarrow \infty
\end{equation}
and
\begin{equation}
\frac{{\tilde f}({\bf k})}{b^3} \sim \frac{8\pi}{(kb)^4}, \qquad k \rightarrow \infty.
\end{equation}
These results are consistent with the general asymptotic result (\ref{decay}).

\begin{figure}
\begin{center}
\includegraphics[  width=2.8in, keepaspectratio,clip=]{fig8a.eps}\vspace{0.2in}
\includegraphics[  width=2.8in, keepaspectratio,clip=]{fig8b.eps} \vspace{0.6in}
\includegraphics[  width=2.4in, keepaspectratio,clip=]{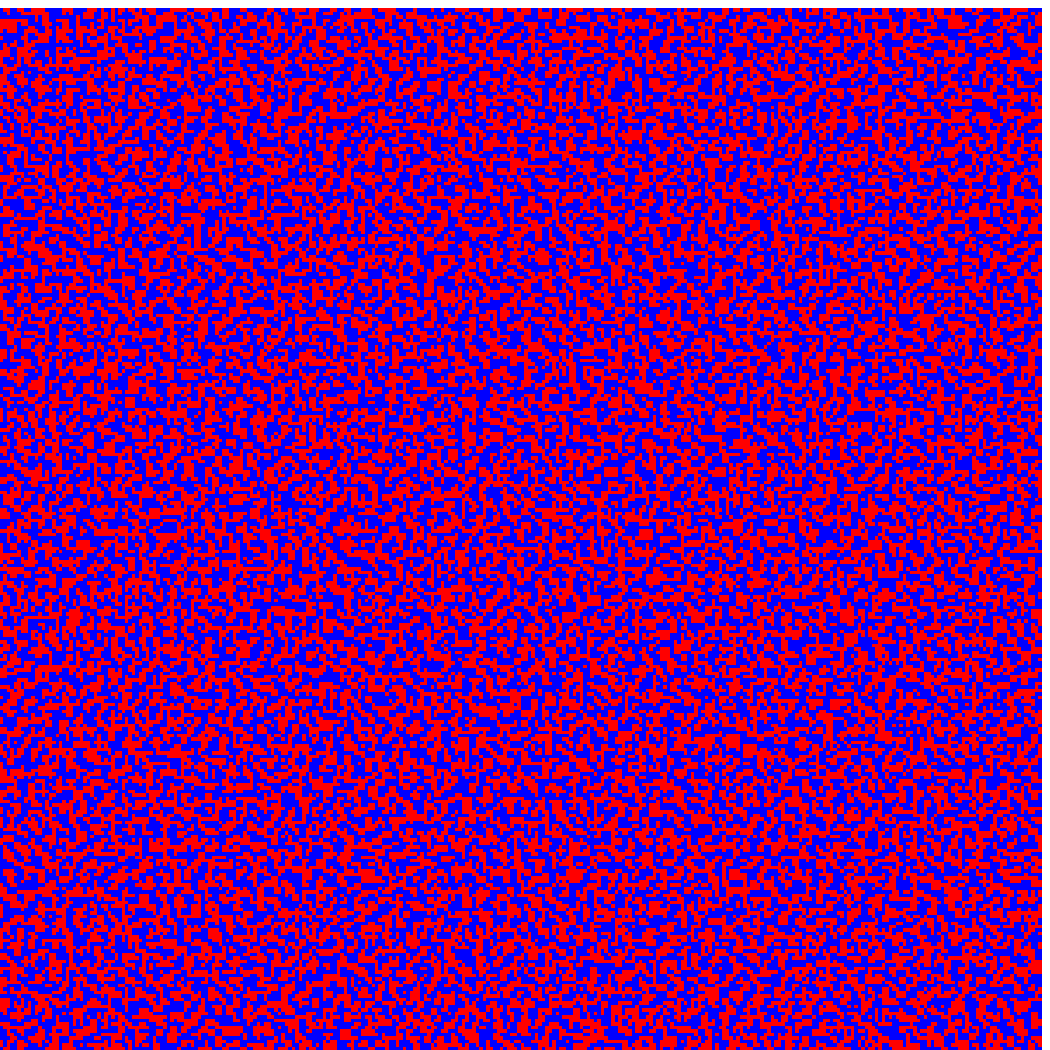} \vspace{-0.4in}
\caption{ Top panel:  Hyperuniform autocovariance function $\chi_{_V}({\bf r})$ given by (\ref{I}) with  $\phi_1=\phi_2=1/2$ in two dimensions where $b=1$, $q=1$ and
in three dimensions where $b=1$ and $q=1/\sqrt{3}$.
Middle panel: Corresponding hyperuniform spectral density ${\tilde \chi}_{_V}({\bf k})$ in two and three dimensions. Bottom panel:  A realization of a disordered hyperuniform two-phase system that corresponds to (\ref{I}) in two dimensions with these parameters, as obtained using reconstruction
techniques \cite{To02a,Ye98a,Ji07}. The final ``energy" is smaller than $10^{-9}$, indicating that the targeted function is achieved
to very high accuracy.}
\label{hypo-1}
\end{center}
\end{figure}

\begin{figure}
\begin{center}
\includegraphics[  width=2.8in, keepaspectratio,clip=]{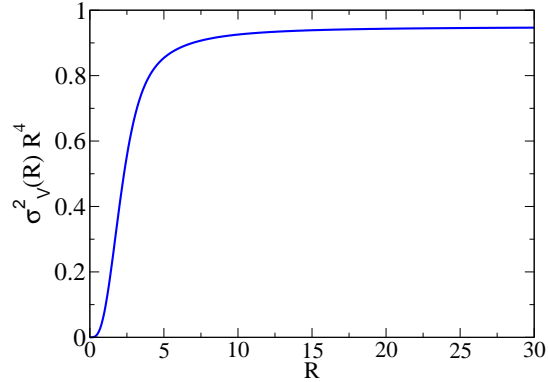}
\caption{Volume-fraction variance $\sigma^2_V(R)$ (multiplied by $R^4$) as a function
of the window radius using (\ref{I}) with  $\phi_1=\phi_2=1/2$ in three dimensions where $b=1$, $q=1/\sqrt{3}$.
The fact that this scaled variance asymptotes to a constant value ($243/256= 0.94921875\ldots$) for large $R$ implies
that this three-dimensional hyperuniform system has a variance that decays like $R^{-4}$, which is 
consistent with the analytical formula (\ref{asy}).}
\label{hypo-var}
\end{center}
\end{figure}

Figure \ref{hypo-1} shows the autocovariance function  and spectral density for a selected set of hyperuniform parameters 
($b$, $q$) in both two and three dimensions.
Not surprisingly, the spectral densities associated with autocovariance functional form (\ref{I}) differ
across dimensions.  To verify that there are indeed
disordered hyperuniform two-phase media that correspond to these autocovariance, well-established
``construction" (reconstruction)  optimization techniques devised by Yeong and Torquato \cite{To02a,Ye98a,Ji07} are employed. Such procedures
utilize simulated-annealing  methods that begin with a random initial guess for a digitized two-phase
system (hypercubic fundamental simulation box that is tessellated into finer hypercubic
cells) satisfying a prescribed volume fraction. The fictitious energy is a  sum of squared
differences between a target correlation function (or corresponding spectral function)
and the correlation function (or corresponding spectral function) of the simulated structure
at any point along the evolution process to the global energy minimum (ground state)
as the fictitious temperature tends to zero. Here we target hyperuniform spectral densities
associated with (\ref{I}).  The bottom panel of Fig. \ref{hypo-1}
shows a final construction in the case of two dimensions
that corresponds to (\ref{I}) with extremely high numerical
accuracy for a selected set of parameters.  Apparently, the known realizability conditions on the function (\ref{I}) are sufficient to 
ensure that it corresponds to a two-phase medium in two dimensions. It is noteworthy that it becomes easier
to ensure realizability of a hypothesized autocovariance function of specific functional form
as the space dimension increases for exactly the same reasons identified for point-configuration
realizability \cite{To06b}. Figure \ref{hypo-var} shows the volume-fraction variance $\sigma^2_{_V}(R)$
as a function of the window radius $R$, as obtained analytically from (\ref{phi-var-1}),  in the case of three dimensions for a selected set
of parameters. We can analytically show that this specific three-dimensional volume-fraction variance has the following
asymptotic scaling:
\begin{equation}
\sigma^2_{_V}(R) \sim \frac{243}{256} \frac{1}{R^4} \qquad (R \rightarrow \infty).
\label{asy}
\end{equation}

\begin{figure}[bthp]
\begin{center}
\includegraphics[  width=2.8in, keepaspectratio,clip=]{fig10a.eps}\vspace{0.2in}
\includegraphics[  width=2.8in, keepaspectratio,clip=]{fig10b.eps} \vspace{0.6in}
\includegraphics[  width=2.4in, keepaspectratio,clip=]{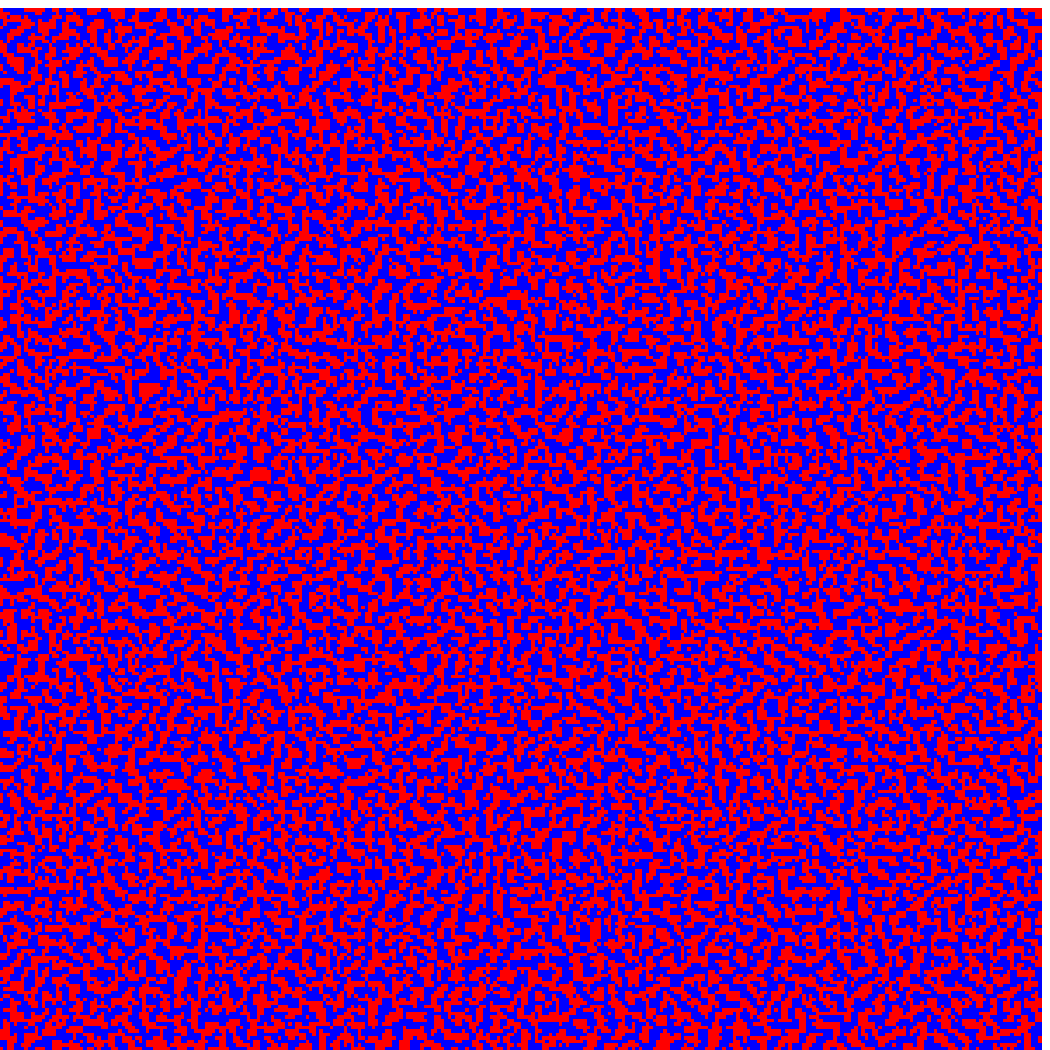} \vspace{-0.4in}
\caption{ Top panel:  Hyperuniform autocovariance function $\chi_{_V}({\bf r})$ given by (\ref{II}) in two dimensions with $a=(1+\sqrt{3})^{3/2}/(\sqrt{2}(3+\sqrt{3}))=0.67479019\ldots$
$b=(\sqrt{6}+\sqrt{2})/2=1.9318516\ldots$, $q=1$ and $\phi_1=\phi_2=1/2$.
Middle panel: Corresponding hyperuniform spectral density ${\tilde \chi}_{_V}({\bf k})$ in two  dimensions. Bottom panel:  A realization of a disordered hyperuniform two-phase system that corresponds to (\ref{II}) in two dimensions with these parameters, as obtained using reconstruction
techniques \cite{To02a,Ye98a,Ji07}. The final ``energy" is smaller than $10^{-9}$, indicating that the targeted function is achieved
to very high accuracy.}
\label{hypo-2}
\end{center}
\end{figure}

As a second example, we consider the function (\ref{case}) in which $\alpha_1=1/2$, $\alpha_2=1/2$ and $\theta=0$, i.e.,
\begin{equation}
f({\bf r})= \frac{1}{2}\exp(-r/a) +\frac{1}{2}\exp(-r/b)\cos(q r),
\label{II}
\end{equation}
which provides greater degrees of freedom to achieve hyperuniformity relative to the form (\ref{I}).
Here  $a$ and $b$ are taken to be positive and thus characteristic length scales.
The specific surface corresponding to (\ref{II}) is given by $s=(a+b)\beta(d)/(2ab\,\phi_1\phi_2)$,
where $\beta(d)$ is the $d$-dimensional constant specified in  (\ref{beta}).
For $d=1$, we find that (\ref{II}) can never correspond to a hyperuniform medium because (\ref{sum-3}) cannot
be satisfied, which also was the case for the function (\ref{I}). This indicates that the hyperuniformity
condition is more difficult to achieve in one dimension than in higher dimensions. 
For $d=2$ and $d=3$, the hyperuniformity sum rule requires that 
\begin{equation}
a= \frac{b((qb)^2-1)^{1/2}}{(qb)^2+1}
\end{equation}
and
\begin{equation}
a= \frac{b(3(qb)^2-1)^{1/3}}{(qb)^2+1}\,
\end{equation}
respectively.
Even though these conditions ensure hyperuniformity in these dimensions, they are not sufficient
to guarantee the  nonnegativity of the spectral density [cf. \ref{1}] for all $k$  because the leading term
in the series expansion of ${\tilde f}({\bf k})$ about $k=0$ is generally  quadratic in $k$ but may have a negative coefficient.
For example, for $d=2$, to ensure positivity of the quadratic term, $qb$ must satisfy the following inequalities:
\begin{equation}
1 < qb \le \frac{1}{2}(\sqrt{6}+\sqrt{2}).
\label{67}
\end{equation}
If $qb$ is equal to  the upper bound in (\ref{67}), the quadratic term vanishes identically such that the leading term 
in the expansion of ${\tilde f}({\bf k})$ about $k=0$ is now quartic in $k$,  which is to be contrasted
with the hyperuniform spectral density associated with (\ref{I}) that goes to zero quadratically in $k$
in the limit $k \rightarrow 0$. 
Under the aforementioned restrictions on the parameters $a$, $b$ and $q$, all of the known realizability conditions described in Sec. \ref{realize} are satisfied.
Figure \ref{hypo-2} shows both the autocovariance function and spectral density for a set of hyperuniform parameters in two dimensions.
The bottom panel of Fig. \ref{hypo-2}
shows a realization obtained by the construction procedure \cite{To02a,Ye98a,Ji07} that corresponds to (\ref{II}) in two dimensions 
with extremely high numerical accuracy for a selected  set of parameters.

\section{Conclusions and Discussion}
\label{con}

For two-phase media in $d$-dimensional Euclidean space $\mathbb{R}^d$ in which one
of the phases  is a packing of spheres, we presented explicit exact expressions for the 
autocovariance function and associated spectral density as well as upper bounds on the volume-fraction 
variance in terms of the number variance for any window radius $R$. We used these results 
to determine the necessary and sufficient conditions for a sphere packing 
to be stealthy and hyperuniform as well as to establish rigorously the requirements
for a  packing comprised of spheres of different sizes to be multihyperuniform.
 We then considered hyperuniformity for general two-phase
media in $\mathbb{R}^d$ outside the class consisting of sphere packings.  We applied realizability conditions for an autocovariance function and its
associated spectral density of a two-phase
medium and incorporated hyperuniformity as a constraint in order to derive new conditions.  We showed that some functional forms
can immediately be eliminated from consideration and identified other forms that are allowable.
Contact was made with well-known two-phase microstructural models (e.g., overlapping spheres and checkerboards) as
well as irregular phase-separation and Turing-type patterns. We ascertained a family of autocovariance functions
that are realizable by disordered hyperuniform two-phase media in arbitrary space dimensions.
Realizations of disordered hyperuniform two-phase media with targeted spectral densities
were explicitly constructed. These studies
elucidate the nature of hyperuniformity in the context of heterogeneous materials.

In a subsequent work, we will explore more fully the explicit construction of disordered hyperuniform
two-phase media and characterize their higher-order statistics (beyond the two-point autocovariance function)
as well as host of other microstructural descriptors that are well-known in homogenization
theory \cite{To02a}. A particularly important goal of such studies will be to develop a deeper understanding
of the  effect of space dimensionality on the microstructural descriptors, including the relevance 
of the ``decorrelation principle" as the space dimension is increased \cite{To06b}.

A fruitful direction for future research would be the study and determination of the effective physical properties
of disordered hyperuniform two-phase systems. There is already evidence demonstrating that disordered hyperuniform cellular
network structures possess novel photonic properties \cite{Man13a,Ha13,Man13b}. However, the investigation
of the bulk properties of general disordered hyperuniform two-phase materials and their technological
relevance is essentially uncharted territory, and its exploration may offer great promise
for novel materials by design. 

Very recently, the hyperuniformity concept was generalized to spin systems and shown
to exist as disordered spin ground states \cite{Ch16a}.
The implications and significance of the existence of such disordered spin systems warrants further study, 
including whether their bulk physical properties, like their many-particle system counterparts, 
are singularly remarkable, and can be experimentally realized. Finally, we note that the notion of
hyperuniformity has recently been generalized to include surface-area fluctuations 
in two-phase media as well as fluctuations associated with random scalar and vector  fields \cite{To16a}.
Now that we know what to look for, different varieties of disordered hyperuniform systems  seem to 
be arising in surprising places and  contexts, and hence offer both intriguing fundamental and applied
research challenges and questions for the future.

\appendix
\section{Fourier Transformation in $d$ Dimensions}
\label{FT}

We employ the following definition of the
Fourier transform of some scalar function $f({\bf r})$ that depends on the 
vector $\bf r$ in  $\mathbb{R}^d$:
\begin{eqnarray}
        \tilde{f}(\mathbf{k}) = \int_{\mathbb{R}^d} f(\mathbf{r}) \exp\left[-i(\mathbf{k}\cdot  \mathbf{r})\right] d\mathbf{r},
\end{eqnarray}
where $\mathbf{k}$ is a wave vector. 
When it is well-defined, the corresponding inverse Fourier transform is given by
\begin{eqnarray}
f(\mathbf{r}) = \left(\frac{1}{2\pi}\right)^d \int_{\mathbb{R}^d}       \tilde{f}(\mathbf{k}) \exp\left[i(\mathbf{k}\cdot  \mathbf{r})\right] d\mathbf{k}.
\end{eqnarray}
If  $f$ is a radial function, i.e., $f$ depends only
on the modulus $r=|\mathbf{r}|$ of the vector $\bf r$, then
its Fourier transform is given by
\begin{eqnarray}
{\tilde f}(k) =\left(2\pi\right)^{\frac{d}{2}}\int_{0}^{\infty}r^{d-1}f(r)
\frac{J_{\left(d/2\right)-1}\!\left(kr\right)}{\left(kr\right)^{\left(d/2\right
)-1}} \,d r,
\label{fourier}
\end{eqnarray}
where  $k=| {\bf k} |$ is wavenumber or modulus of the wave vector $\bf k$
and $J_{\nu}(x)$ is the Bessel function of order $\nu$.
The inverse transform of $\tilde{f}(k)$ is given by
\begin{eqnarray}
f(r) =\frac{1}{\left(2\pi\right)^{\frac{d}{2}}}\int_{0}^{\infty}k^{d-1}\tilde{f}(k)
\frac{J_{\left(d/2\right)-1}\!\left(kr\right)}{\left(kr\right)^{\left(d/2\right
)-1}} d k.
\label{inverse}
\end{eqnarray}
We recall the first several terms in  the series expansion of $J_{\nu}(x)$ 
about $x=0$:
\begin{equation}
\hspace{-0.6in}J_{\nu}(x) =  \frac{(x/2)^{\nu}}{\Gamma(\nu +1)}- \frac{(x/2)^{\nu+2}}{\Gamma(\nu +2)}+ \frac{(x/2)^{\nu+4}}{2\Gamma(\nu +3)} -\frac{(x/2)^{\nu+6}}{6\Gamma(\nu +4)} +{\cal O}(x^{\nu +8}),
\end{equation}
which we apply in Sec. \ref{two}.

\section{Packings of Spheres with a Size Distribution and Multihyperuniformity}
\label{size}

Both the autocovariance and associated spectral density for packings of hard spheres with a 
continuous or discrete size distribution at overall number density $\rho$ have been
derived \cite{To02a,Lu91}. We collect these results here and apply them to 
establish rigorously the requirements for multihyperuniformity \cite{Ji14}.

In the case of a continuous distribution in radius ${\cal R}$ characterized by
a probability density function $f({\cal R})$ that normalizes to unity, 
\begin{equation}
\int_{0}^{\infty} f({\cal R}) d{\cal R}=1,
\label{f-normal}
\end{equation}
the packing fraction and the autocovariance function  are given respectively by \cite{Lu91,To02a}
\begin{equation}
\phi = \rho \int_{0}^{\infty} f({\cal R}) v_1(R) d{\cal R}
\label{pack}
\end{equation}
and 
\begin{eqnarray}
\hspace{-0.2in}\chi_{_V}({\bf r})& =&  \rho \int_{0}^{\infty} f({\cal R}) v_2^{int}(r;{\cal R}) d{\cal R} \nonumber\\
&+&\rho^2 \hspace{-0.05in} \int_{0}^{\infty} \hspace{-0.13in} d{\cal R}_1  \hspace{-0.07in}  \int_{0}^{\infty}  \hspace{-0.13in}  d{\cal R}_2  f({\cal R}_1)  f({\cal R}_2)  
 \,m(r;{\cal R}_1) \otimes m(r;{\cal R}_2) \otimes h({\bf r};{\cal R}_1,{\cal R}_2), \nonumber\\
\label{chi-poly-r}
\end{eqnarray}
where $h({\bf r};{\cal R}_1,{\cal R}_2)$ is the appropriate generalization
of the total correlation function for the centers of two spheres of radii ${\cal R}_1$ and ${\cal R}_2$ separated
by a distance $r$. Fourier transformation of (\ref{chi-poly-r}) gives the corresponding
spectral density
\begin{eqnarray}
\hspace{-0.2in}{\tilde \chi}_{_V}({\bf k}) & =&  \rho \int_{0}^{\infty} f({\cal R}) {\tilde m}^2(k;{\cal R}) d{\cal R} \nonumber\\
&+&\rho^2 \hspace{-0.05in} \int_{0}^{\infty} \hspace{-0.13in} d{\cal R}_1  \hspace{-0.07in}  \int_{0}^{\infty}  \hspace{-0.13in}  d{\cal R}_2  f({\cal R}_1)  f({\cal R}_2)  
 \,{\tilde m}(k;{\cal R}_1)  {\tilde m}(k;{\cal R}_2)  {\tilde h}({\bf k};{\cal R}_1,{\cal R}_2).
\label{chi-poly-k}
\end{eqnarray}

One can obtain corresponding results for  spheres
with $M$ different  radii $a_1,a_2,\ldots,a_M$ from the continuous case by letting
\begin{equation}
f(R) = \sum_{i=1}^{M} \frac{\rho_{i}}{\rho} \delta ({\cal R} -a_{i}), 
\label{f-discrete}
\end{equation}
where $\rho_{i}$ is the number density 
of type-$i$ particles, respectively, and $\rho$ is the {\it total number density}.
Substitution of (\ref{f-discrete}) into (\ref{pack}), (\ref{chi-poly-r}) and (\ref{chi-poly-k}) yields 
the corresponding packing fraction, autocovariance function and spectral density, respectively, as
\begin{equation}
\phi =\sum_{i=1}^M \rho_i v_1(a_i),
\end{equation}
\begin{equation}
\hspace{-0.4in}\chi_{_V}({\bf r})  = \sum_{i=1}^M \rho_i v_2^{int}(r;a_i)  
+  \sum_{i=1}^M  \sum_{j=1}^M \rho_i \rho_j  
 \,m(r;a_i) \otimes m(r;a_j) \otimes h({\bf r};a_i,a_j)
\label{chi-poly-r-dis}
\end{equation}
and
\begin{equation}
\hspace{-0.4in}{\tilde \chi}_{_V}({\bf k}) =\sum_{i=1}^M \rho_i {\tilde m}^2(k;a_i)  S({\bf k};a_i)
+  \sum_{i\neq j}^M   \rho_i \rho_j  
 \,{\tilde m}(k;a_i) {\tilde m}(k;a_j) {\tilde h}({\bf k};a_i,a_j),
\label{chi-poly-k-dis}
\end{equation}
where 
\begin{equation}
S({\bf k};a_i)\equiv 1+\rho_i {\tilde h}({\bf k};a_i,a_i)
\end{equation}
is the structure factor for  type-$i$ particles.
It immediately follows that the spectral density at the origin
is given by
\begin{eqnarray}
{\tilde \chi}_{_V}(0) &=&\sum_{i=1}^M \rho_i v_1^2(a_i)  S(0;a_i) \nonumber\\
&+&  \sum_{i\neq j}^M   \rho_i \rho_j  
 \,v_1(a_i) v_1(a_j) {\tilde h}(0;a_i,a_j).
\label{chi-dis-k-hyp}
\end{eqnarray}

We now prove that when each subpacking associated with each component is hyperuniform,
i.e., the first term on the right side of (\ref{chi-dis-k-hyp})
is zero, the second term must also be identically zero (sum of cross terms vanish), leading to the hyperuniformity
of the entire packing, i.e.,  ${\tilde \chi}_{_V}(0)=0$. Such a polydisperse
packing has been called multihyperuniform \cite{Ji14}. 

We begin the proof by considering the spectral density ${\tilde \chi}_{_V}({\bf k})$ for a general, very large but 
finite-sized two-phase heterogeneous system that is contained within 
hypercubic fundamental cell in $\mathbb{R}^d$ of side length $L$ and volume $V=L^d$ subjected
to  periodic boundary conditions, which is given by \cite{To99c}:
\begin{equation}
{\tilde  \chi}_{_V}({\bf k})= \frac{|{\tilde {\cal J}}({\bf k})|^2}{V}, \qquad {\bf k} \neq {\bf 0},
\label{discrete}
\end{equation}
where ${\tilde {\cal J}}({\bf k})$ is the discrete Fourier transform of  phase indicator function
minus the phase volume fraction, which generally is a complex number for any $\bf k \neq 0$.
Ultimately, we take the thermodynamic  limit to make contact with (\ref{chi-dis-k-hyp}). 
Consider the two-phase system to be a finite packing of $N$ spheres consisting of
$M$ components. Let ${\bf r}_1^{(i)},\ldots,  {\bf r}^{(i)}_{N_i}$ denote the positions
of type-$i$ spheres, where $N_i$ is the total number of spheres of radius $a_i$
and $i=1,2,\ldots,M$ such that $N=\sum_{i=1}^M N_i$. The 
discrete Fourier representation of the ``scattering amplitude" ${\tilde {\cal J}}({\bf k})$
for such a multicomponent packing
was given in Ref.  \cite{Za11c}, which can be recast as follows: 
\begin{equation}
{\tilde {\cal J}}({\bf k})= \sum_{i=1}^M {\tilde m}(k;a_i) f({\bf k};a_i),
\label{J}
\end{equation}
where
\begin{equation}
f({\bf k};a_i)= \sum_{n=1}^{N_i} \exp(i{\bf k}\cdot {\bf r}^{(i)}_n),
\end{equation}
and the product $ {\tilde m}(k;a_i) f({\bf k};a_i)$ represents the scattering amplitude
for the $i$th component (subpacking). Substitution of (\ref{J}) into (\ref{discrete}) yields
\begin{equation}
\hspace{-0.6in}{\tilde \chi}_{_V}({\bf k}) =\sum_{i=1}^M \rho_i {\tilde m}^2(k;a_i)  {\cal S}({\bf k};a_i)
+  \sum_{i\neq j}^M   \rho_i \rho_j  
 \,{\tilde m}(k;a_i) {\tilde m}(k;a_j) {\cal H}({\bf k};a_i,a_j),
\label{c}
\end{equation}
where
\begin{equation}
{\cal S}({\bf k};a_i)= \frac{|f({\bf k};a_i)|^2}{N_i}
\end{equation}
is the discrete structure factor for the $i$th subpacking and
\begin{equation}
{\cal H}({\bf k};a_i,a_j)= \frac{f({\bf k};a_i)f({\bf k};a_j)\,V}{N_iN_j}.
\end{equation}
The similarity between this discrete representation of the spectral density
of a multicomponent packing and the continuous version (\ref{chi-poly-k-dis})
is readily apparent.

The chosen hypercubic fundamental cell restricts the wave vectors to take  discrete
values, which  are defined by the vectors
that span the reciprocal hypercubic lattice, i.e., ${\bf k}=(2\pi n_1/L,2\pi n_2/L, \cdots, 2\pi n_d/L)$,
where $n_i$ ($i=1,2,\ldots,d$) are the integers. Thus, the smallest positive wave vectors have
magnitude  $k_{min}=2\pi/L$. Now we constrain the scattering amplitude for each
component to be zero at $|{\bf k}|=k_{min}$, 
which from (\ref{J}) implies that  ${\tilde {\cal J}}(|{\bf k}|=k_{min})=0$ and hence
${\tilde \chi}_{_V}(|{\bf k}|= k_{min})=0$. This in turn means that the second sum (\ref{c}) involving
the cross terms must vanish at  $|{\bf k}|= k_{min}$.  To complete the proof, we must
consider the thermodynamic limit because the ensemble-average relation (\ref{chi-dis-k-hyp}) 
applies under this condition. Assuming ergodic media,  the zero-wave-vector behavior
of the spectral density  defined by (\ref{chi-dis-k-hyp}) can be extracted from
the discrete spectrum  (\ref{discrete}) in the thermodynamic limit, i.e.,
\begin{equation}
{\tilde \chi}_{_V}({\bf 0}) \equiv \lim_{N, V\rightarrow \infty} {\tilde \chi}_{_V}(|{\bf k}|= k_{{min}}).
\end{equation}
The limit here is taken at constant number density $\rho = N/V$, implying that 
$k_{min} \rightarrow 0$. Under the prescribed conditions mentioned above, we
find that ${\tilde \chi}_{_V}({\bf 0})=0$, which completes the proof.

\ack

The author dedicates this article to the honor the memory of 
Professor George Stell, mentor and colleague, for his seminal
research contributions to statistical mechanics.
The author thanks  Duyu Chen, Jaeuk Kim and Zheng Ma 
for their careful reading of the manuscript. He is especially grateful to Duyu Chen
for his assistance in  creating the microstructure-construction figure. This work was supported by the 
National Science Foundation under Grant No. DMS-1211087.

%\bibliographystyle{unsrt}
%\bibliography{new}

\end{document}